\documentclass[aps,prd,amsmath,floats,floatfix,notitlepage,twocolumn,
superscriptaddress,nofootinbib,showpacs,longbibliography]{revtex4-1}

\usepackage[english]{babel}
\usepackage[utf8]{inputenc}
\usepackage[colorinlistoftodos, color=green!40, prependcaption]{todonotes}
\usepackage[pdftex, pdftitle={Article}, pdfauthor={Author}]{hyperref} 
\usepackage{graphicx} 
\usepackage{microtype}
\usepackage{titlesec}
\usepackage{calrsfs}
\DeclareMathAlphabet{\pazocal}{OMS}{zplm}{m}{n}
\usepackage{comment}
\usepackage{physics,xfrac}
\usepackage{subcaption}
\usepackage{amsmath,amssymb,xcolor}
\usepackage{cancel}
\usepackage{tikz}
\usepackage{caption}
\usepackage{ragged2e}
\usepackage{orcidlink}

\usepackage[letterpaper,top=2.5cm,bottom=2.5cm,left=1.7cm,right=1.7cm,marginparwidth=2cm]{geometry}
\newcommand{\enigma}{\texttt{ENIGMA}}

\newcommand{\esigmahm}{\texttt{ESIGMAHM}}

\newcommand{\seobnrvfourphm}{\texttt{SEOBNRv4PHM}}

\newcommand{\nrsurdqfour}{\texttt{NRSur7dq4}}

\usepackage[english]{babel}
\usepackage[normalem]{ulem}
\definecolor{darkgreen}{rgb}{0,0.5,0}

\hypersetup{
    unicode=false,          
    pdftoolbar=true,        
    pdfmenubar=true,        
    pdfstartview={FitH},    
    pdftitle={IMRESIGMA},    
    colorlinks=true,       
    linkcolor=red,          
    citecolor=blue,        
    filecolor=magenta,      
    urlcolor=darkgreen,           
    linktocpage=true
}

\begin{document}

\newcommand{\Cardiff}{\affiliation{Gravity Exploration Institute, School of Physics and Astronomy,
Cardiff University, Cardiff, CF24 3AA, United Kingdom}} %
\newcommand{\CSGC}{\affiliation{Centre for Strings, Gravitation and Cosmology, Department of Physics, Indian Institute of Technology Madras, Chennai 600036, India}} %
\newcommand{\AEIP}{\affiliation{Max Planck Institute for Gravitational Physics
    (Albert Einstein Institute), Am M{\"u}hlenberg 1, 14476 Potsdam, Germany}}
\newcommand{\AEIH}{\affiliation{Max Planck Institute for Gravitational Physics
    (Albert Einstein Institute), Callinstra{\ss}e 38, D-30167 Hannover, Germany}}
\newcommand{\IITM}{\affiliation{Department of Physics, Indian Institute of Technology Madras, Chennai 600036, India}} %
\newcommand{\ICTS}{\affiliation{
    International Centre for Theoretical Sciences,
    Tata Institute of Fundamental Research, Bangalore 560089, India
}}
\newcommand{\BITSG}{\affiliation{
    Department of Physics, BITS-Pilani, K K Birla Goa Campus, Zuarinagar, Goa 403726, India
}}
\newcommand{\CIFAR}{\affiliation{
    Canadian Institute for Advanced Research, CIFAR Azrieli Global Scholar, MaRS Centre, West Tower, 661 University Ave, Suite 505, Toronto, ON M5G
 1M1, Canada
}}

\title{Early Warning From Eccentric Compact Binaries:\\ Template Initialization And Sub-dominant Mode Effects
}

\author{Priyanka Sinha \orcidlink{0009-0004-1300-6775}}\email[]{priyanka.sinha@icts.res.in} \ICTS{}
\author{R. Prasad \orcidlink{0000-0002-6602-3913}}\ICTS{}
\author{Mukesh Kumar Singh \orcidlink{0000-0001-8081-4888}}\Cardiff{}\ICTS{}
\author{Prayush Kumar \orcidlink{0000-0001-5523-4603}}\ICTS{}
\author{Akash Maurya \orcidlink{0009-0006-9399-9168}}\ICTS{}
\author{Kaushik Paul \orcidlink{0000-0002-8406-6503}}\IITM{}\CSGC{}

\date{\today}

\begin{abstract}
Early warning of gravitational waves (GWs) is essential for multi-messenger observations of binary neutron star and black hole-neutron star merger events. In this study, we investigate early warning prospects from eccentric compact binaries, whose mergers are expected to comprise a significant fraction of detected GW events in the future. Eccentric binaries exhibit oscillatory frequency evolution, causing GW frequencies to recur multiple times through their coalescence. Consequently, generating eccentric waveform templates for early warning requires specification of initial conditions. While the standard approach involves initiating waveform generation when the orbit-averaged frequency enters the detector band, we compare this with an alternative approach that uses the periastron frequency as the starting point. Our analysis shows that initializing at the periastron frequency yields an improved signal-to-noise ratio and sky localization. Additionally, including subdominant modes alongside the dominant $(2,2)$ mode leads to further improvements in sky localization. We explore the parameter space of primary mass $m_1 \in [1.4, 15] \, M_\odot$, spin $\chi_1 \in [0, 0.8]$, and eccentricity $e \leq 0.4$ across three detector configurations: O5, Voyager, and 3G. We find that in the O5 configuration, including eccentricity and subdominant modes, the sky localization area can be reduced by 2–80 \% at 1000 square degrees (from $e_5 = 0.1$ to $e_5 = 0.4$), offering up to $41$ seconds of additional early warning time. For NSBH systems, subdominant modes alone contribute up to $70 \%$ reduction. For the Voyager configuration, the sky area reduction due to eccentricity spans $2-85 \%$, with eccentricity increasing from 0.1 to 0.4, yielding up to 1 minute of extra early warning time at 1000 sq. deg. sky area. Subdominant modes contribute up to $94 \%$ reduction for NSBH systems, though their impact is nearly zero for BNS systems. In the 3G detector scenario, the sky area reduction due to eccentricity reaches 80 \% (from $e_{2.5} = 0.1$ to $e_{2.5} = 0.4$) at 100 square degrees, and subdominant modes enhance the reduction up to $98 \%$ for NSBH systems.
\end{abstract}

\maketitle

\pagebreak

\section{Introduction}
\label{sec:introduction}

The LIGO-Virgo-KAGRA detector network \cite{LIGOScientific:2014pky, VIRGO:2014yos, KAGRA:2020tym} has made significant strides, completing three observational runs (O1, O2, O3) \cite{abbott2023gwtc, LIGOScientific:2018mvr, LIGOScientific:2020ibl} and is currently in its fourth run. Notably, O1 marked the historic detection of the first binary black hole merger  \cite{abbott2016gw150914}, whereas O2 observed the first binary neutron star merger (BNS) and its electromagnetic signals \cite{abbott2017gw170817}, and O3 reported the first neutron star-black hole (NSBH) merger \cite{LIGOScientific:2021qlt}. These events, particularly the BNS and NSBH systems, have gained significant interest due to their potential to constrain the neutron star equation of state \cite{abbott2018gw170817, annala2018gravitational, dietrich2020multimessenger}, facilitate independent measurements of the Hubble constant \cite{LIGOScientific:2017adf, chen2018two}, uncover the origins of short gamma-ray bursts \cite{giacomazzo2012compact, foucart2020brief, gottlieb2023unified}, and provide insights into heavy element nucleosynthesis during mergers \cite{just2015comprehensive,  metzger2017kilonovae, rosswog2024heavy}. 

The GW170817 event, observed through multimessenger astronomy, has already highlighted the significant science that can be gained from such events \cite{abbott2017multi, abbott2019tests, shappee2017early, abbott2018gw170817, ligo2018gw170817, creminelli2017dark, ezquiaga2017dark}. The advanced LIGO and Virgo detectors detected this event with a combined signal-to-noise ratio (SNR) of $32$. The inspiral phase lasted about 100 seconds within the detectors' sensitive band, followed by a gamma-ray burst 1.7 seconds later \cite{goldstein2017ordinary,savchenko2017integral}, which could be associated with the event. The combined data from three GW detectors enabled an initial sky localization of 31 square degrees \cite{abbott2017gw170817}. An electromagnetic follow-up campaign successfully identified a counterpart near the galaxy NGC4993 \cite{coulter2017swope, valenti2017discovery, tanvir2013kilonova, lipunov2017master, arcavi2017optical, abbott2017multi}, consistent with the localization and distance estimates derived from the GW signals. Early ultraviolet (UV) monitoring revealed a blue transient that faded within 48 hours \cite{evans2017swift,shappee2017early}, while optical and infrared data indicated a redward evolution over 10 days \cite{abbott2017multi, kasliwal2022spitzer}. Radio and X-ray emissions were initially absent and emerged approximately 9 and 16 days after the merger \cite{troja2017x, hallinan2017radio}. These observations allowed us to study the merger remnant, its environment, and the distinct physical processes at play. Another BNS event, GW190425, was observed but poorly localized initially~\cite{abbott2020gw190425}.
This event, however, did not have an associated gamma-ray burst nor any electromagnetic signals associated with it \cite{coughlin2019growth,paterson2021searches,boersma2021search, coulter2024gravity}.
Three notable NSBH merger events have also been observed: GW190426, GW200105, and GW200115 \cite{abbott2021observation}. For GW200105, the absence of data from the LIGO Hanford detector led to poor sky localization. In contrast, GW200115 was observed by multiple detectors and localized to an area of approximately 900 square degrees \cite{abbott2021observation}. Although these events may have involved partial tidal disruption of the neutron star, potentially producing an accretion disk and ejecta \cite{paschalidis2017general, foucart2020brief}, no associated electromagnetic counterparts were detected despite extensive follow-up observations \cite{hosseinzadeh2019follow,lundquist2019searches,gompertz2020searching,paterson2021searches,oates2021swift,goldstein2019growth,gourdji2023lofar,anand2021optical,dichiara2021constraints}.
{\it Truly multi-messenger events are therefore relatively rare, but rich in scientific information. These observations underscore the importance of early warning systems and precise sky localization in facilitating successful follow-up of those rare BNS and NSBH mergers which will be accompanied by multi-messenger signals}~\cite{lazzati2017off,lazzati2018late}.

Compact-object binaries formed through isolated stellar evolution are expected to radiate away most of their orbital eccentricity by the time they enter the GW detector's sensitivity band. In contrast, those formed through dynamical interactions in dense environments, such as galactic centers and globular clusters, can retain significant eccentricity when they become detectable~\cite{Postnov:2014tza,mapelli2020binary,park2017black, grobner2020binary, ford2022binary}. Additionally, hierarchical triple systems can provide another pathway to eccentric mergers via the Kozai-Lidov mechanism~\cite{antognini2014rapid,petrovich2017greatly}. Therefore, identifying orbital eccentricity serves as a key tool for distinguishing between different binary formation channels. 
The majority of GW detections to date are consistent with quasi-circular binary orbits~\cite{LIGOScientific:2018mvr,LIGOScientific:2020ibl,abbott2023gwtc,Antonini:2022vib}. Some independent studies have suggested the presence of eccentricity, however, there is little consensus on any specific event~\cite{romero2022four, gupte2024evidence, planas2025eccentric, romero2021signs, Gayathri:2020coq, Morras:2025xfu}.
With third-generation GW observatories, like the Einstein Telescope (ET)~\cite{punturo2010einstein,Hild_2012} and Cosmic Explorer (CE)~\cite{LIGOScientific:2016wof,reitze2019cosmic}, we can explore vast regions of our universe \cite{bailes2021gravitational,borhanian2024listening}, bringing numerous galactic centers, globular clusters, and other potential sites into observational focus. Eccentric mergers involving BNS and NSBH systems could constitute a sizable fraction of the detected events \cite{lee2010short, hoang2020neutron,dhurkunde2023search}. The presence of eccentricity in the last inspiral stages can lead to complex and intriguing dynamical outcomes, especially in cases with high eccentricity \cite{stephens2011eccentric, east2012eccentric, east2015eccentric, zenati2024mass}. {\it To maximize the scientific gains of these relatively rare eccentric merger events jointly with EM observations, robust early warning is even more essential.} 
Early warning and localization of eccentric merger events would allow telescopes to capture the earliest possible glimpse of the merger, enabling the observation of potential precursors \cite{tsang2012resonant}, the intermediate merger products such as the remnant disk or hypermassive neutron stars \cite{hotokezaka2013remnant}, or track the detailed evolution of the kilonova light curves \cite{douchin2001unified,arcavi2018first,shappee2017early,cowperthwaite2017electromagnetic}.

The possibility of early warnings for quasi-circular compact binaries has been extensively studied, including the effect of subdominant modes in addition to the dominant $\ell=|m|=2$ mode of gravitational radiation~\cite{kapadia2020harbingers, singh2021improved, singh2022improved}. {Similarly, early warning for spin-precessing NSBH binaries has also been investigated for near-future detectors \cite{Tsutsui:2021izf}.}
A key relationship governing the frequency of these emitted GWs in non-precessing binary orbits is $f_{\ell m}(t) \simeq m f_{orb}(t)$ \cite{creighton2012gravitational}, where $f_{\ell m}(t)$ represents the instantaneous frequency of waveform multipoles with mode numbers $(\ell,m)$, and $f_{orb}(t)$ is the orbital frequency. This relationship directly implies that at any specific moment, certain subdominant modes oscillate at frequencies higher than that of the dominant mode. As a consequence, they enter the sensitive frequency band of GW detectors earlier than the dominant mode. And their time spent within this detector band is significantly longer, specifically by a factor of $(m/2)^{8/3}$ compared to the quadrupole modes~\cite{sathyaprakash1994filtering}.
This offers a crucial advantage:
by including these modes in the analysis, we can achieve a more extended early warning time for binary mergers compared to relying solely on the dominant mode.

For eccentric binary systems, prior studies \cite{kyutoku2014pre,yang2024eccentricity, yang2024advantage} have investigated the potential for early warning using higher harmonics induced by eccentricity in the dominant $\ell = |m| = 2$ waveform multipole. These studies utilized a non-spinning, inspiral eccentric waveform model in frequency domain, constructed by applying the stationary phase approximation to the post-circular (PC) expansion of eccentric time-domain inspirals~\cite{yunes2009post,huerta2014accurate}. 
These eccentricity-induced PC harmonics can indeed oscillate at higher frequencies
than the equivalent quasi-circular (zero-eccentricity) harmonic, allowing them to enter the GW detector's frequency band earlier.
Refs. \cite{yang2024eccentricity, yang2024advantage} utilized these PC harmonics to demonstrate the potential for early warning for eccentric signals. In addition, recent work has shown that for signals with non-negligible eccentricity in detector band, an eccentricity-optimized algorithm can significantly improve the sky localization areas~\cite{Pal:2025iyo}, laying the foundation for a matched-filtering based eccentric early-warning system. 

In this study, we build upon existing works and study the impact of waveform modes sourced by higher-order source multipoles than the quadrupole. Specifically, while Refs. \cite{yang2024eccentricity, yang2024advantage} include eccentricity-induced harmonics of the quadrupolar GW emission alone, we include eccentric GW modes sourced by octupolar and higher-order source multipoles by including $h_{\ell m}$ with $\ell \geq 2$. {\it Doing so allows us to estimate the maximum possible early warning possibilities for eccentric spinning binary mergers.} See Figs.~\ref{fig:modes_plot_eccentric_time_domain} and~\ref{fig:hm_contribution_eccentric_gws} for the relative contributions of these sub-dominant waveform modes.
Our work utilizes the time-domain eccentric inspiral-merger-ringdown (IMR) waveform model, \esigmahm{} \cite{paul2024esigmahm}. This model incorporate the effects of component spins (aligned to the orbital angular momentum) and orbital eccentricity. It supports mild to moderate initial orbital eccentricities as its merger portion uses a numerical relativity (NR) surrogate model restricted to quasi-circular orbits. Our early warning calculations are based on the methodology outlined in Refs. \cite{kapadia2020harbingers, singh2021improved, singh2022improved}, but we have extended this approach to accommodate eccentric GW signals. 
We study the impact of orbital eccentricity, subdominant modes, and the primary component's parameters on sky localization of both neutron star black hole (NSBH) binaries and binary neutron stars (BNS). Our findings indicate that orbital eccentricity has a significant effect, enabling more precise sky localization than non-eccentric (quasi-circular) binaries, with higher eccentricities providing greater improvements. For instance, in the BNS binaries in the O5 observing scenario, an eccentricity of $e_5 = 0.4$ can lead to a $40\%$ reduction in sky-localization area compared to quasi-circular BNS (using only the dominant quadrupolar modes). This improvement is found during the early-warning interval when the eccentric binary has been localized to $100-1000$ square degrees. In similar scenarios, eccentric NSBH binaries see up to a $80\%$ reduction in sky-localization area. When subdominant modes are included in the analysis, eccentric systems exhibit even more substantial improvements: up to $85 \%$ reduction in sky area was found for eccentric NSBH sources, accompanied by $27 \%$ increase in early warning time. 
We find that both mass ratio and orbital eccentricity comparably enhance the contribution of sub-dominant waveform harmonics, with eccentricity playing a more pronounced effect at lower mass ratios. In contrast, the influence of BH spin on both sky localization and early warning is found to be relatively weak. These gains further scale with detector sensitivity: Voyager offers greater improvements than O5, and 3G outperforms Voyager. {\it From these results we conclude that it is imperative to include sub-dominant harmonics when attempting low-latency sky-localization to provide early-warning for potentially EM-bright BNS and NSBH merger events.}
We detail these findings in Sec.~\ref{subsec: higher_mode_contribution_eccentric}.

Furthermore, analyzing eccentric sources in Fourier domain using waveform templates introduces a complexity regarding the initial conditions of said templates for Bayesian likelihood integration. For quasicircular binaries one can generate a unique waveform from a specific orbital frequency onward; but any given frequency occurs multiple times for eccentric inspirals, implying that the starting point of template generation for likelihood integration play a subtle role. Unlike quasi-circular binaries, whose orbital frequency increases monotonically, the instantaneous frequency of eccentric binary signals oscillates significantly during each orbit. 
Conventionally one generates waveform templates when the orbit-averaged frequency enters the frequency band of the detector, given initial values for orbital eccentricity and the mean anomaly angle that measures the angular position of both components in an eccentric orbit. However there remain some cycles in this moment's past when the instantaneous orbital (and therefore GW) frequency was already in the detector band. This was clearly pointed out in Fig.~3 of Ref.~\cite{Shaikh:2023ypz}, and we illustrate it in Fig.~\ref{fig:frequency_evolution} for further discussion. Ignoring these extra cycles limits the sky localization of the source to be suboptimal. This motivates the need to instead specify the initial conditions of analysis templates by fixing the mean anomaly angle to the periastron (since that is where the instantaneous orbital frequency is highest during an orbit), and specifying the initial orbital eccentricity and periastron frequency.

Currently, template banks used in early-warning efforts~\cite{Sachdev:2020lfd,magee2021first} typically fix a reference orbital/GW frequency and grid over the binary's intrinsic parameters, such as component masses and spin angular momentum, as defined at the specified orbital/GW frequency~\cite{owen1996search, babak2006template, ajith2008template}. While this approach is sufficient for quasi-circular binaries, the development of robust early-warning template banks for eccentric binaries might require gridding templates over periastron frequencies and initial eccentricity instead. In this paper we study both of these options. 
We show that using periastron frequencies and eccentricity as initial conditions yields measurably better sky localization compared to the alternative. By the time a fiducial NSBH binary ($m_1=10M_\odot, m_2=1.4M_\odot$) with eccentricity $e=0.3$ has been localized to a fiducial sky area of, say, $10^3 \, \text{sq. deg.}$, we would gained an additional early warning time of around $7.8\,\text{s}$, $13.5\,\text{s}$, and $185\,\text{s}$ in the O5, Voyager, and 3G scenarios, respectively by starting our waveform filters from periastron frequency. This gain will be an important improvement in current early-warning systems already, and will be especially significant for the third-generation detector networks ~\cite{Sachdev:2020lfd,magee2021first,kovalam2022early}. 
{\it We therefore conclude that the choice between these two ways of specifying the initial conditions of analyses templates for eccentric merger events can have a significant impact on the effectiveness of early-warning sky localization. Early-warning template banks for eccentric binary merger events should grid templates over periastron frequencies and initial eccentricities to optimize their efficacy.} Note that this applies generally to both time-domain and frequency-domain template models. We detail our results in Sec.~\ref{subsec: initial_conditions_for_eccentric_templates}.

The article is organized as follows. Sec.~\ref{sec:methods} describes methods for early warning and sky localization, various observing scenarios considered, and the eccentric waveform model used in this paper. Sec.~\ref{sec:results} presents the results, with a particular focus on the initial conditions for waveform and SNR integration for eccentric GW signals, as well as the effects of subdominant modes on sky localization and early warning time across the parameter space of mass, spin, and eccentricity in different observing scenarios. Finally, Section \ref{sec:conclusion} offers a summary and outlook for future research on the subject.

\section{Methods}
\label{sec:methods} 

\subsection{Compact binaries \& subdominant modes}
\label{subsec:Compact_binaries_higher_modes} 

Gravitational waves from coalescing compact binaries with component masses $m_1$ and $m_2$ at luminosity distance $d_L$ are detected as a real-valued time series $h(t)$. This signal $h(t)$ represents the response of the detector to the plus ($h_{+}$) and cross ($h_{\times}$) polarizations of the GW. The measured strain is given by \cite{Sathyaprakash:2009xs}, 
\begin{align}\label{eq:h_of_t}
  h(t) = F_+ h_+(t) + F_{\times} h_{\times}(t),  
\end{align}

where $F_+$ and $F_\times$ are the response functions of GW detectors, which depend on the direction of propagation of the GW with respect to the plane of the detector. The two polarizations above can be decomposed in the basis of spin-weighted spherical harmonics as \cite{newman1966note}
\begin{align}
    h_+ - i h_\times = \frac{1}{d_L} \sum_{\ell=2}^{\infty} \sum_{m=-l}^{\ell} h_{\ell m}(t, \vec{\lambda}) Y_{\ell m}^{-2}(\iota, \varphi_0),
\end{align}

where $h_{\ell m}$ are coefficients of this multipolar expansion, or simply, multipoles of the waveform, and ${}^{-2}Y_{\ell m}$ are spin $-2$ weighted spherical harmonics \cite{thorne1980multipole}. $\vec{\lambda}$ includes all intrinsic binary parameters such as component masses, spins, {orbital eccentricity}, etc; and $\iota$ and $\phi_0$ are polar and azimuthal angles from the source centered frame. For non-precessing binaries, the positive and negative $m$ multipoles are related as~\cite{kidder2008using} $h_{\ell,-m} = (-1)^\ell \, h_{\ell, m}^*$.

\begin{figure}[h]
    \centering
    \includegraphics[width=1\linewidth]{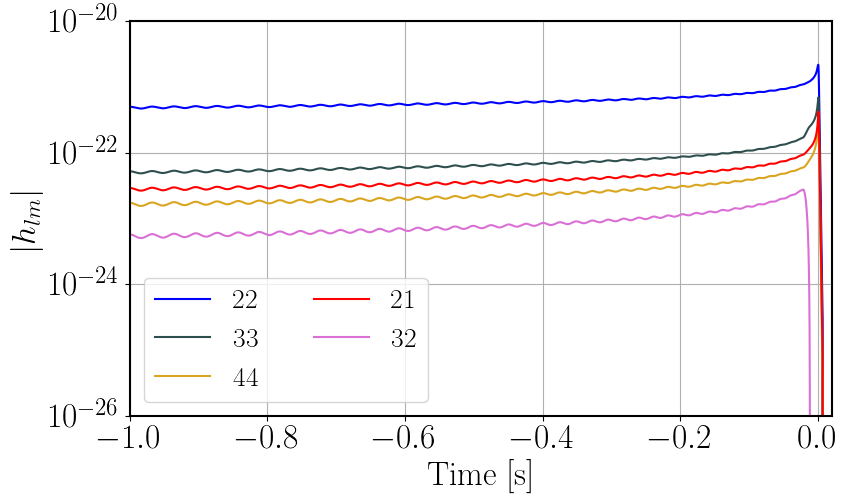}
    \caption{\justifying Amplitude of the dominant and subdominant modes of GWs for an eccentric binary with $m_1$=10$M_\odot$, $m_2$= 1.4$M_\odot$, $S_{1z}= 0.25$, $S_{2z}= 0.0$, $e_{10}$= 0.2 and $d_L=40$ Mpc.
    }
    \label{fig:modes_plot_eccentric_time_domain}
\end{figure}

\begin{figure*}
\centering
\begin{subfigure}{.32\textwidth}
    \centering
    \includegraphics[width=\textwidth]{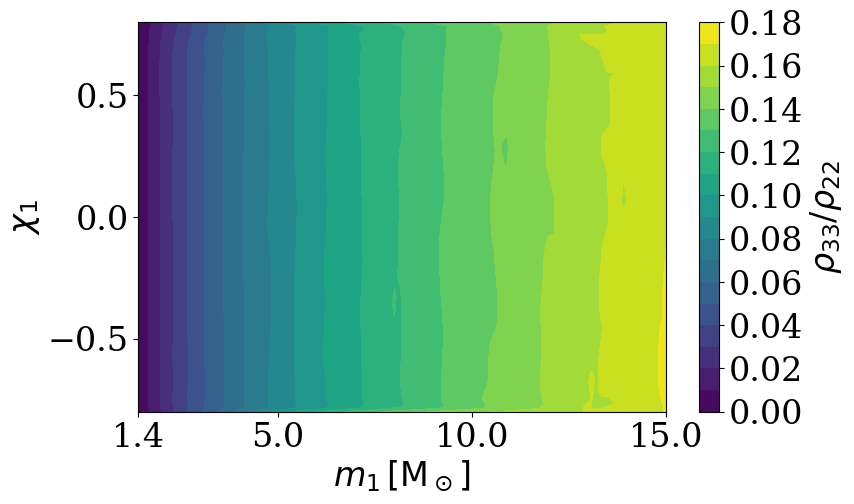} 
    \label{SUBFIGURE LABEL 1}
\end{subfigure}
\begin{subfigure}{.32\textwidth}
    \centering
    \includegraphics[width=\textwidth]{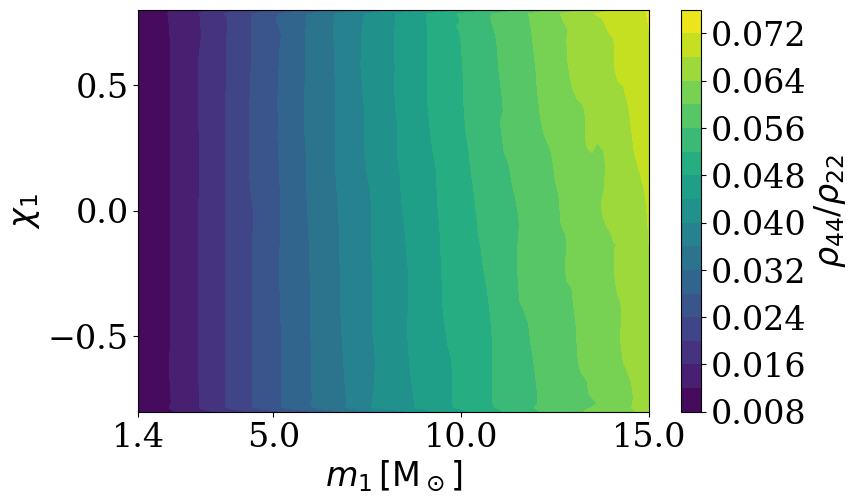}  
    \label{SUBFIGURE LABEL 4}
\end{subfigure}
\begin{subfigure}{.32\textwidth}
    \centering
    \includegraphics[width=\textwidth]{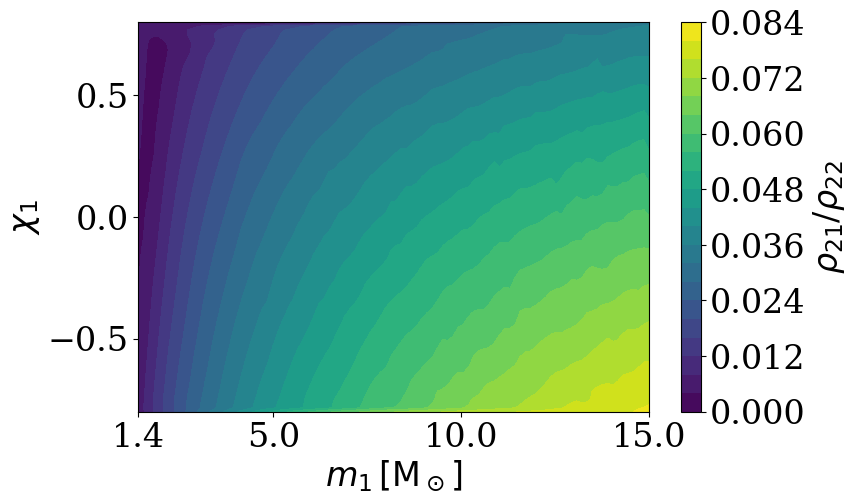} 
    \label{SUBFIGURE LABEL 1}
\end{subfigure}
\begin{subfigure}{.32\textwidth}
    \centering
    \includegraphics[width=\textwidth]{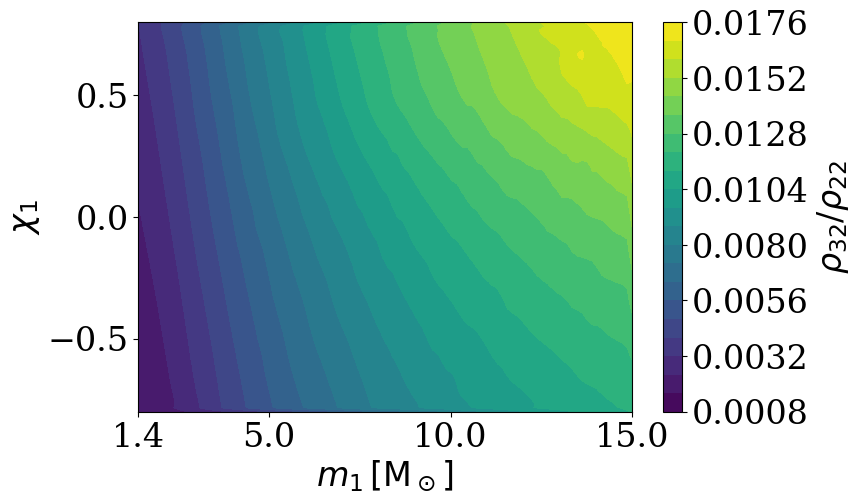}  
    \label{SUBFIGURE LABEL 4}
\end{subfigure}
\begin{subfigure}{.32\textwidth}
    \centering
    \includegraphics[width=\textwidth]{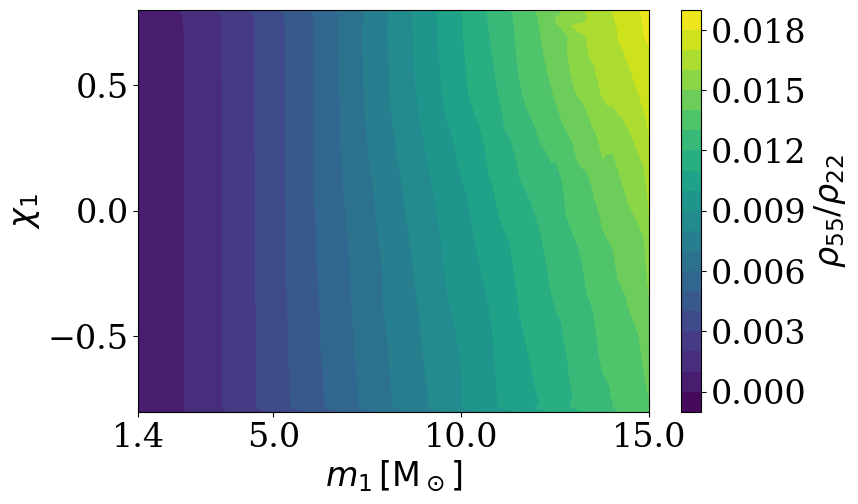}  
    \label{SUBFIGURE LABEL 1}
\end{subfigure}
\begin{subfigure}{.32\textwidth}
    \centering
    \includegraphics[width=\textwidth]{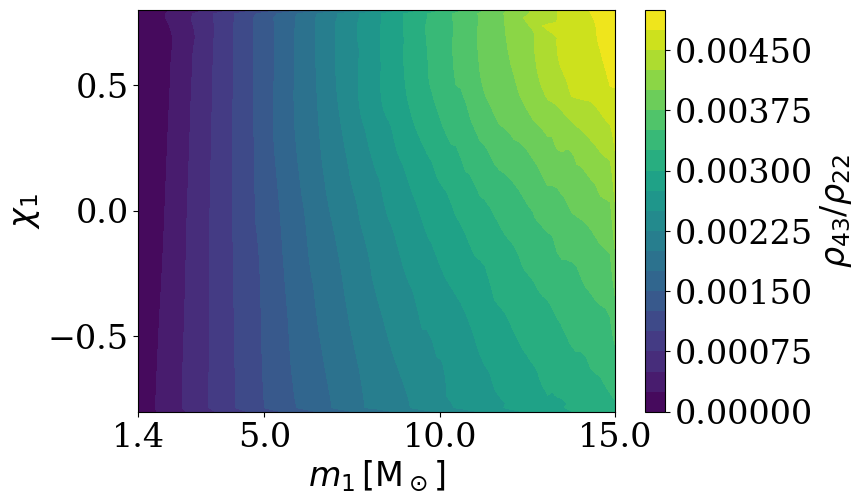}  
    \label{SUBFIGURE LABEL 4}
\end{subfigure}
\caption{\justifying SNR ratio of subdominant mode relative to dominant (2,2) mode for parameter space of mass $m_1$ and dimensionless spin $\chi_1= [0, 0, \chi_{1z}]$ of primary object, while fixing secondary objects mass $m_2$= 1.4 $M_\odot$, $\chi_2= [0, 0, 0]$. Eccentricity at 10 Hz frequency $e_{10} = 0.2$, note that  $e_{10}$ denotes the orbital eccentricity at the instant when the orbit-averaged frequency of the dominant $h_{2,2}$ crosses $10$Hz.}

\label{fig:hm_contribution_eccentric_gws}
\end{figure*} 

The dominant contribution of the GW signals comes from the quadrupolar emission with $\ell=|m|=2$~\cite{peters1963gravitational,peters1964gravitational}. The next subdominant multipoles are $(\ell=3,m=\pm3)$, $(\ell=4,m=\pm4)$, $(\ell=5,m=\pm5)$, and $(\ell=2,m=\pm1)$~\cite{kapadia2020harbingers} for quasi-circular non-spinning GW signals, and beyond these the contribution becomes small. For systems with eccentricity and aligned spins, we investigate and find that the mode hierarchy remains unchanged. In Fig. \ref{fig:modes_plot_eccentric_time_domain}, the amplitude of different modes for a fiducial NSBH with $m_1= 10M_\odot$, $m_2= 1.4M_\odot$, $\chi_{1z}= 0.25$, $\chi_{2z}= 0.0$ and $e_{10}= 0.2$ is shown. It is evident that the amplitude of subdominant modes is lower but may contribute modestly relative to (2,2) mode.

We next analyzed the relative contribution of subdominant modes for a parameter space spanned by $m_1\in [1.4, 15] M_\odot$, $\chi_1\in [-0.8, 0.8]$, while fixing the secondary component mass $m_2=1.4 M_\odot$, and its spin $\chi_2= 0.0$. We also fix the orbital eccentricity $e_{10}= 0.2$. In our notation, $e_{10}$ denotes the orbital eccentricity at the instant when the orbit-averaged frequency of the dominant $h_{2,2}$ crosses $10$Hz. We consider multipoles with: $(\ell, |m|) = (2,2), (2,1), (3,3), (3,2), (4,4), (4,3), (5,5)$. We calculated the SNR $\rho_{\ell m}$ for all modes,
\begin{align}
    \rho_{\ell m}^2 = 4 \int_{f_{low}}^{f_{high}} \frac{|\tilde{h}_{\ell m}(f)|^2}{S_n(|f|)} df,
\end{align}
where $\tilde{h}_{\ell m}(f)$ is the Fourier transform of the mode $h_{\ell m}$, and $S_n(|f|)$ is the design power spectral density (PSD) for LIGO \cite{LIGO:T0900288}, setting the frequency range from $f_{low}= 10 \, \text{Hz}$ to $f_{high}= 1000 \, \text{Hz}$ to see the contribution of HM relative to the dominant mode over the parameter space. 
In Fig. \ref{fig:hm_contribution_eccentric_gws}, we present the SNR ratio for different subdominant modes versus the dominant mode: $\rho_{\ell m}/\rho_{22}$. It can be seen that at mass ratio ($q=m_1/m_2$) of 10, (3, 3) mode is contributing around 17$\%$, for (4, 4) and (2, 1) mode, this is around 7$\%$, other modes have relative contributions less than 2$\%$. As we increase the mass ratio, the HM contribution increases as they get excited due to the mass asymmetry of the system; with increasing spin, the SNR contribution does not vary much for $(3, 3), (4, 4), (5, 5)$ mode but does for $(2, 1), (3, 2), (4, 3)$ mode, we can see some variation with their overall contribution being higher for binaries with anti-aligned primary spin. This is likely due to the merger frequency for these anti-aligned-spin configurations being reduced enough to lie close to the most sensitive frequency range for the GW detectors. Furthermore, we observe that increasing the eccentricity to $e_{10}= 0.4$ minimally changes the relative contribution from subdominant modes, compared to the $e_{10}=0.2$ scenario. For BNS systems, the contribution of the $(3,3), (4,4), (2,1)$ modes is observed to be below 2\%, while other subdominant modes contribute less than 1\%. Based on these results, in the remainder of this paper we include $(\ell, |m|)=(2,1),(2,2),(3,3),(4,4)$ modes to study the sky localization capabilities for both NSBH and BNS sources, and discuss the results in Sec.~\ref{subsec: higher_mode_contribution_eccentric}.

\subsection{Localizing GW sources}\label{subsec:localizing_gw_sources}

Practical low-latency GW astronomy efforts, such as GstLAL \cite{cannon2021gstlal, sachdev2019gstlal}, typically employ matched-filtering techniques across a network of detectors to identify signals and subsequently perform sky localization using rapid parameter estimation algorithms. Early warning from binary neutron stars has been shown using GstLAL and SPIIR searches \cite{Sachdev:2020lfd, magee2021first, kovalam2022early}. Other works have also proposed machine-learning based systems for early warning~\cite{Wei:2020xrl}. Our goal here, however, is to understand the theoretical limits on how well we can localize a given source with and without subdominant modes. For this purpose, we adopt the Fisher-matrix based approach outlined in Refs.~\cite{fairhurst2009triangulation, fairhurst2011source} and summarize it below. 

For a network of detector, sky localization uncertainty can be computed using information in the time-of-arrival measurement uncertainty for GW signals in each detector, and their separation. Let us define frequency moments of GW waveform, assuming stationary and Gaussian noise in GW detectors, as
\begin{align}
    \bar{f}^n \equiv 4 \int_{0}^{\infty} f^n \frac{|\tilde{h}(f)|^2}{S(f)} df,
\end{align}
where $\tilde{h}(f)$ is the Fourier transform of $h(t)$ defined in Eq.~\ref{eq:h_of_t}.
The SNR and the effective bandwidth can be computed from these frequency moments as:
\begin{align}
    \rho^2 = \Bar{f}^{ 0}; \, \, \, \, \sigma_f^2 = \frac{\Bar{f}^2}{\rho^2} - \left(\frac{\Bar{f}^1}{\rho^2} \right)^2.
\end{align}
From these two, we can calculate the expected uncertainty in the arrival time of the GW signal at the detector,
\begin{align}
   \sigma_t = \frac{1}{2 \pi \rho \sigma_f}. 
\end{align}
Let $\mathbf{R}$ be the source's actual sky location in an Earth centered coordinate frame (making it a unit vector), and $T_0$ be the actual time the GW signal passes through the Earth's center. Then the time at which the GW signal passes through the $i$-th detector is $T_i = T_0 - \mathbf{R}\cdot\mathbf{d_i}$;
given $d_i$ is the separation between detector $i$ and the center of Earth. Similarly we can write the measured time at which the GW signal passes through the $i$-th detector, given the measured sky position of the source $\mathbf{r}$, as $t_i = t_0 -\mathbf{r}\cdot\mathbf{d_i}$
The probability distribution of measured arrival time of GW signal in various detectors given the actual arrival time can be written using the timing uncertainties for all detectors ($\sigma_i$):
\begin{align}
   p(\vec{t}|\vec{T}) = \prod_i \frac{1}{\sqrt{2\pi}\sigma_i} \exp{\frac{-(t_i - T_i)^2}{2\sigma_i^2}},
\end{align}
where $\vec{T}$ and $\vec{t}$ are the vectors of the {\it actual} and {\it measured} time of arrivals at all the detector in the network. Applying Bayes' theorem, the posterior distribution of the arrival time given the prior distribution $ p(T_i)$ is,
\begin{align}
    p(\vec{T}|\vec{t}) \propto p(T_i) \exp{-\sum_i \frac{(t_i - T_i)^2}{2\sigma_i^2}}.
\end{align}
Using the relation between source position and arrival time of GW at different detectors, we can express $T_i$ and $t_i$ in terms of $\mathbf{R}$ and $\mathbf{r}$. Then, the posterior distribution for actual source position $\mathbf{R}$ given the measured source position $\mathbf{r}$ is (a detailed derivation of this provided in the Appendix. of \cite{fairhurst2011source}),
\begin{align}\label{eq:p_R_r}
   p(\mathbf{R}|\mathbf{r}, t_0) \propto p(\mathbf{R}) \exp{-\frac{1}{2} (\mathbf{r}-\mathbf{R})^T \mathbf{M} (\mathbf{r}-\mathbf{R})}, 
\end{align}
where $\mathbf{M}$ is the Fisher Matrix that depends on the timing uncertainties $\sigma_i$ and the separation $\mathbf{D}_{i,j} = \mathbf{d}_i - \mathbf{d}_j$ between detectors $i$ and $j$. It is given by
\begin{align}
    \mathbf{M} = \frac{1}{\sum_k \sigma_k^{-2}} \sum_{i,j} \frac{\mathbf{D}_{i,j} \mathbf{D}_{i,j}^{T}}{2\sigma_i^2 \sigma_j^2}.
\end{align}
The sky-localization area of the source with a probability $p$ can be estimated by projecting the Fisher Matrix $\mathbf{M}$ onto the directions orthogonal to $\mathbf{r}$ and then using the two smallest eigenvalues of the projected matrix. If $\sigma_1$ and $\sigma_2$ are the localization accuracies of the eigen-directions of the projected matrix, then the calculated sky area for the source with probability $p$ is \cite{fairhurst2011source},
\begin{align}\label{eq:area}
    \mathrm{Area}(p) \approx 2\pi \sigma_1 \sigma_2 [-\ln{(1-p)}].
\end{align}
{In our sky area calculations, we use Eq.~\eqref{eq:area} for various network scenarios (detailed in \ref{subsec: observing_scenario}) to localize BNS/NSBH sources with a probability of $p=0.9$.}

\subsection{Source Models}\label{subsec:Source_Models}
In this paper we use the model \esigmahm{} \cite{paul2024esigmahm}, a time-domain Eccentric Spinning Inspiral Generalized Merger Approximant with Higher-order Modes which extends \enigma{} \cite{Huerta:2016rwp,huerta2018eccentric,chen2021observation} by incorporating the effects of aligned component spins and including multiple subdominant waveform harmonics.  \esigmahm{} combines an eccentric inspiral piece with a quasi-circular merger-ringdown (MR) stitched together by matching and aligning the phase during a {\it transition} time-window. The inspiral evolution of \esigmahm{} combines results from post-Newtonian theory~\cite{blanchet2014gravitational}, self force approach~\cite{Poisson:2011nh}, and black hole perturbation theory~\cite{Sasaki:2003xr}, and the MR piece is based on a surrogate model for quasicircular numerical relativity (NR) simulations. \esigmahm{} requires that the binary nearly circularize by the time the model transitions from inspiral to MR prescription.
It provides subdominant waveform modes $h_{\ell m}$ with $(\ell, m) = (2,\pm1) , (3,\pm 3), (3,\pm 2), (4, \pm 4), (4,\pm 3)$ along with dominant ($2,\pm2$) multipoles.

\esigmahm{} models the inspiral dynamics on eccentric orbits using quasi-Keplerian variables. The conservative dynamics is encoded in the time evolution dynamics of PN mean anomaly ($l$) and instantaneous phase ($\phi$):
\begin{align}
    M \Dot{\phi}= x^{3/2} \left[ \sum_{i=0}^{3} c_i x^i + \order{x^4} \right],
\end{align}
\begin{align}
    M\Dot{l}= x^{3/2} \left[1+ \sum_{i=1}^{3} d_i x^i + \order{x^4}\right].
\end{align}
And the radiative dynamics in the evolution equations of PN parameter $x$ and orbital eccentricity $e$, both encoding the orbital energy and angular momentum radiated away,
\begin{align}
    M\Dot{x}= x^5 \left[\sum_{i=0}^4 y_i x^i + \order{x^{9/2}}\right],
\end{align}
\begin{align}
    M\Dot{e}= x^4 \left[\sum_{i=0}^3 z_i x^i + \order{x^4}\right].
\end{align}
From these, the orbital separation ($r$) and eccentric anomaly ($u$) can be obtained by solving the following coupled equations which map $(l, x)\rightarrow(r, u)$:
\begin{align}
    \frac{r}{M}= \frac{1-e\cos{u}}{x} + \sum_{i=1}^{3} a_i x^{i-1},
\end{align}
\begin{align}
    l= u - e\sin{u} + \sum_{i=2}^{3} b_i x^i.
\end{align}
The coefficients $\{c_i, d_i, y_i, z_i\}$ can be found in Refs.~\cite{Hinder:2008kv, Huerta:2016rwp,Henry:2023tka, paul2024esigmahm}.
The spherical harmonic modes of the waveform are computed using relations that include both instantaneous and hereditary effects. 
The instantaneous terms in waveform modes include non-spinning and spinning corrections for general orbits~\cite{Mishra:2015bqa, Paul:2022xfy, Henry:2023tka}, while the hereditary part includes corrections that are available in the zero-eccentricity limit~\citep{Henry:2022dzx}.
See table 1 of \cite{paul2024esigmahm} for a summary of all the terms that are included in both the equations of motion and waveform mode expressions.
The merger-ringdown uses the numerical relativity surrogate model \nrsurdqfour{}~\cite{Varma:2019csw} capable of creating short waveforms for spinning precessing binaries on quasi-spherical orbits with high accuracy. This choice restricts \esigmahm{} to model eccentric binaries that nearly circularize in the final stages of inspiral. As the domain of \nrsurdqfour{} is restricted along the mass-ratio dimension, we substitute it for \seobnrvfourphm{} to model higher mass-ratio mergers. 
In this work we use the code implementation of \esigmahm{} in the project \texttt{esigmapy}~\cite{esigmapy}.

\subsection{Observing scenarios}\label{subsec: observing_scenario}

\begin{table*}[]
    \centering
    \caption{Initial parameters for GW waveform generation and SNR integration for different observing scenarios}
    {\renewcommand{\arraystretch}{1.5}
    \begin{tabular}{|p{3cm}|p{2cm}|p{2cm}|p{2cm}|p{3cm}|p{3.5cm}|}
    \hline
     Observing scenario & Waveform starting\, \,frequency & Integration starting \, \, frequency & Eccentricity defined at & Modes for Sec.~\ref{subsec: initial_conditions_for_eccentric_templates} & Modes for Sec.~\ref{subsec: higher_mode_contribution_eccentric} \\
    \hline
     O5 & $5$ Hz & $10$ Hz & $5$ Hz & $(2,2)$ & $(2,2), (3,3), (4,4), (2,1)$ \\ 
     Voyager & $5$ Hz & $10$ Hz & $5$ Hz & $(2,2)$  & $(2,2), (3,3), (4,4), (2,1)$ \\ 
     3G (ET, $2\times$CE) & $2.5$ Hz & $5$ Hz & $2.5$ Hz & $(2,2)$ & $(2,2), (3,3), (4,4), (2,1)$ \\ 
    \hline
    \end{tabular}}
    \label{tab:table}
\end{table*}

\begin{figure*}
    \centering
    \begin{subfigure}[b]{0.48\textwidth}
         \centering
         \includegraphics[width=\textwidth]{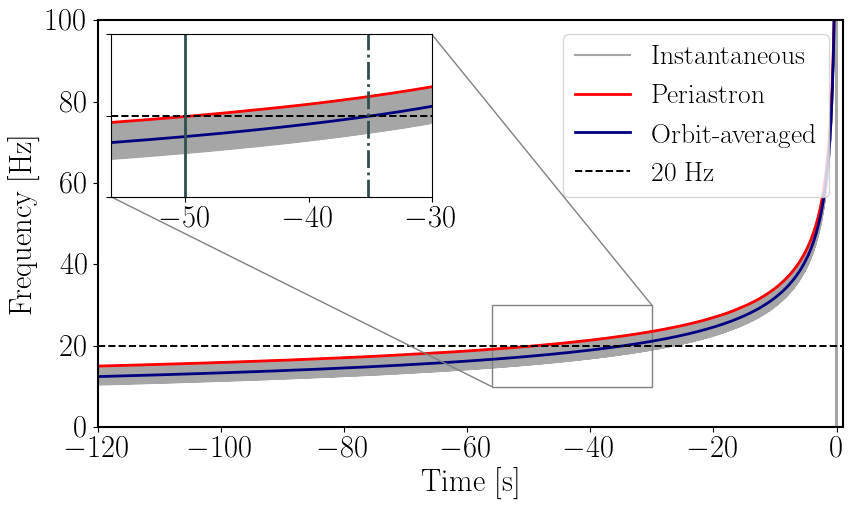}
         
    \end{subfigure}
    \hfill
     \begin{subfigure}[b]{0.48\textwidth}
         \centering
         \includegraphics[width=\textwidth]{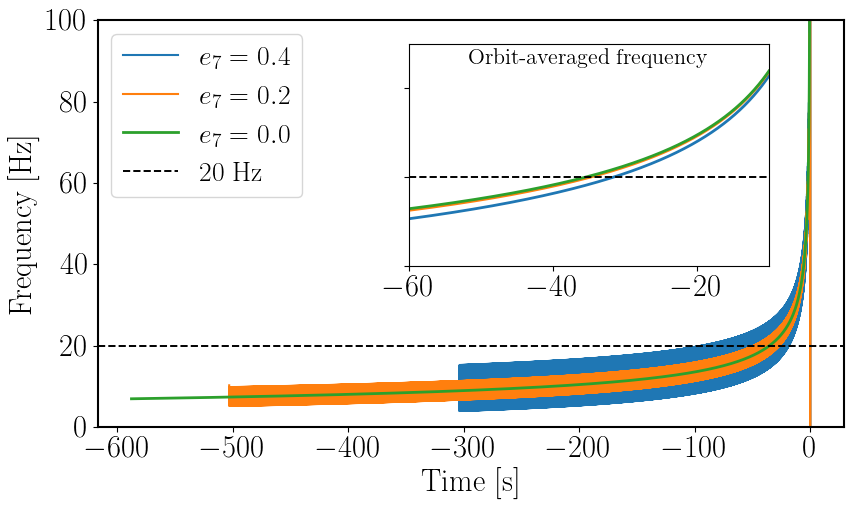}
         
    \end{subfigure}
    \caption{\justifying Left Panel: Frequency evolution of the $(2,2)$ mode of the GW signal emitted by an eccentric binary system. The components of this binary have masses $m_1 = 10 \, M_\odot$ and $m_2 = 1.4 \, M_\odot$, with an initial eccentricity of $e_7 = 0.2$ at a frequency of 7 Hz. The gray curve shows the instantaneous GW frequency, {it appears almost as a band because we show an extended time span spanning hundreds of orbits}; the red curve indicates the interpolated periastron frequency; and the blue curve denotes the orbit-averaged frequency evolution. The inset provides a zoomed-in view near 20 Hz. In this inset, the solid vertical line marks the point at which the periastron frequency reaches 20 Hz, while the dash-dotted line marks the crossing of the orbit-averaged frequency. The time difference between these two events is approximately 15 seconds. Right Panel: Frequency evolution for different values of initial eccentricity. As eccentricity increases, the instantaneous frequency enters the frequency band of interest earlier. However, the inset reveals that the orbit-averaged frequency crosses the band later for signals originating from more eccentric systems. 
    }
    \label{fig:frequency_evolution}
\end{figure*}

In this work, we consider three imminent observing scenarios of the ground-based GW detector network: O5, Voyager, and 3G.
The O5 scenario consists of a network of 5 detectors (LIGO Livingston, LIGO Hanford, LIGO-India, Virgo, and KAGRA). The three LIGO detectors are assumed to have the same sensitivity as A+ \cite{LIGOAplusDesign, LIGO2018, Saleem:2021iwi}, while Virgo and KAGRA's sensitivity is taken to be the same as Virgo's sensitivity corresponding to a BNS range of 260 Mpc \cite{abbott2020prospects, KAGRA2019}.
In our Voyager scenario, the sensitivity of the three LIGO detectors in the O5 configuration is upgraded to Voyager sensitivity~\cite{LIGO2015, adhikari2019astrophysical}, while that of Virgo and KAGRA are upgraded to A+ sensitivity~\cite{LIGOAplusDesign}. For the 3G scenario, we use a network of $3$ detectors: two Cosmic Explorer (CE)~\cite{LIGOScientific:2016wof,reitze2019cosmic}
detectors with sensitivity corresponding to a BNS range of $4200$ Mpc, and one Einstein Telescope (ET)~\cite{punturo2010einstein,Hild_2012} with sensitivity as in \cite{abbott2017exploring, LIGO2018}. The geographical coordinates of both CE detectors are identical to that of LIGO Livingston and Hanford, while that of ET is kept similar to Virgo's. For optimal sky-location in the O5 scenario~\footnote{{The optimal sky-location here corresponds to the point in the sky where a detector network has maximal sensitivity}}, we choose right ascension (RA), declination (DEC), polarization angle value as $60.06^\circ$, $3.99^\circ$, $331.41^\circ$ respectively~\cite{kapadia2020harbingers}. For the 3G scenario the same angles are $197.45^\circ$, $113.38^\circ$, $0^\circ$ respectively~\cite{kapadia2020harbingers}. Also we consider inclination angle ($\iota$) to be $60^\circ$ to enhance the contribution of subdominant modes. In Table~\ref{tab:table} we present the initial parameters for all the different scenarios. Throughout this article, the BNS and NSBH binaries considered in the analysis are assumed to be at a luminosity distance of $d_L=40$ Mpc.

\section{Results}\label{sec:results}

\subsection{Initial conditions for Eccentric GW waveform templates}\label{subsec: initial_conditions_for_eccentric_templates}

Analyses of GW detector data are performed most efficiently in Fourier domain, as that makes it computationally inexpensive compared to doing so in time domain, and also the noise power spectral density is most naturally defined and measured in frequency domain. Analyses of eccentric binary coalescence signals using waveform templates for Bayesian likelihood integration can get complicated by the choice of template initial conditions. 
For quasi-circular binaries, the orbital frequency increases monotonically allowing for unique waveform generation from a given starting frequency, which is often the lower frequency cutoff of the sensitive frequency band of GW detectors. The instantaneous frequency of eccentric systems, on the other hand, oscillates during every orbit. Consequently, any specific frequency value can be crossed multiple times, rendering the template's starting point critical for accurate parameter estimation.
Conventionally, templates are initialized when the orbit-averaged frequency enters the detector's band, given an initial eccentricity ($e_0$) and mean anomaly ($l_0$). However, this approach neglects preceding orbital cycles where the instantaneous gravitational-wave (GW) frequency may have already been within the sensitive band~\cite{Shaikh:2023ypz}. The omission of these early, in-band cycles can lead to suboptimal source sky localization.

We illustrate this in Fig.~\ref{fig:frequency_evolution}. In its left panel we show the frequency evolution of dominant $(2,2)$ mode of GW signal from an eccentric NSBH binary with $m_1=10 \, \mathrm{M_\odot}, m_2= 1.4 \, 
\mathrm{M_\odot}$, and $e_{7}= 0.2$, located at a distance of $40$ Mpc. Different colors represent different frequency evolutions: the instantaneous frequency evolution is shown in gray, the orbit-averaged in blue, and the frequency of periastron points in red. We observe that the periastron frequency enters the frequency band (starting at $20$ Hz) at around $-50$ seconds, whereas the orbit-averaged frequency enters the $20$ Hz frequency at around $-35$ seconds, and the time delay between these two is $15$ seconds. For a binary with $e_7= 0.4$, this time delay increases to nearly $60$ seconds. So, the GW cycles associated with periastron passages enter the sensitive band much earlier than when the orbit-averaged frequency does. Ignoring these early contributions can result in loss of SNR and reduced early warning time. 
The right panel of Fig.~\ref{fig:frequency_evolution} illustrates the frequency evolution of the dominant mode emitted by eccentric binary systems with $m_1=10 \, \mathrm{M_\odot}$, $m_2=1.4 \, \mathrm{M_\odot}$, for different eccentricity values. Specifically, we considered eccentricities $e_7 \in \{0, 0.2, 0.4\}$. The instantaneous frequency evolutions reveal that signals from more eccentric binaries oscillate at higher frequencies and enter the detection band significantly earlier, despite their shorter overall duration. Conversely, the inset displays the orbit-averaged frequencies for the same signals. The orbit-averaged frequency of eccentric signals is consistently lower than that of quasi-circular signals, leading to their delayed entry into the GW detectors' sensitive band.

To address this, we propose an alternative specification for template initial conditions: fixing the mean anomaly to its value at periastron (where instantaneous orbital frequency is maximal) and defining the template by an initial eccentricity and instantaneous frequency.
Current template banks for early-warning systems~\cite{Sachdev:2020lfd,magee2021first} are typically constructed by fixing a reference frequency and gridding over intrinsic parameters (e.g., masses, spins) defined at that frequency~\cite{owen1996search,babak2006template,ajith2008template}. While adequate for quasi-circular systems, robust early-warning template banks for eccentric binaries may necessitate gridding over parameters such as periastron frequency and initial eccentricity. In this section we investigate the implications of both of these template initialization strategies in the context of early-warning capabilities with current and future ground-based GW detector network. We compare them by quantifying the resulting loss in sky-localization information from the suboptimal strategy, across a range of initial eccentricities for different observation scenarios.

\begin{figure*}
     \centering
     \begin{subfigure}[b]{0.48\textwidth}
         \centering
         \includegraphics[width=\textwidth]{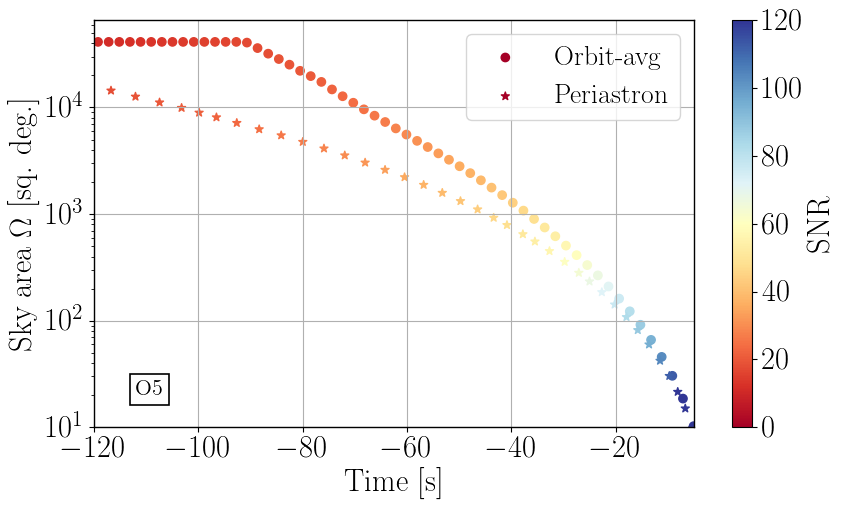}
     \end{subfigure}
     \hfill
     \begin{subfigure}[b]{0.48\textwidth}
         \centering
         \includegraphics[width=\textwidth]{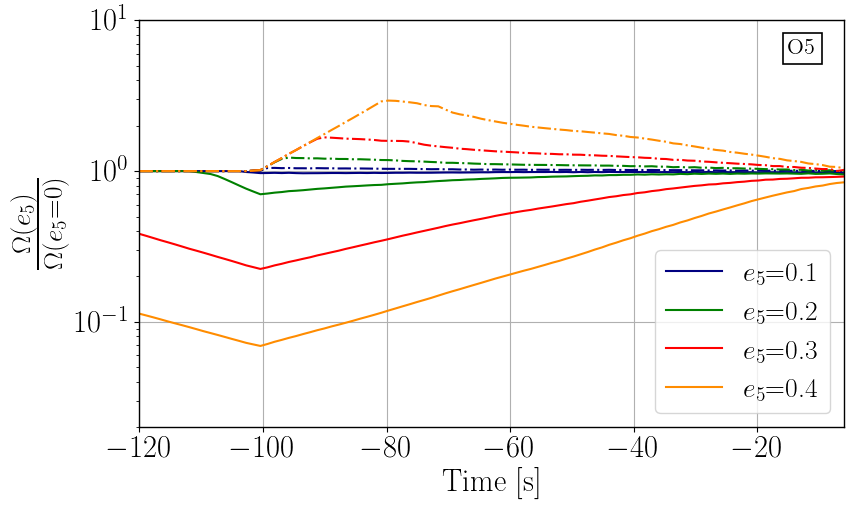}
         
    \end{subfigure}
    \hfill
    \begin{subfigure}[b]{0.48\textwidth}
         \centering
         \includegraphics[width=\textwidth]{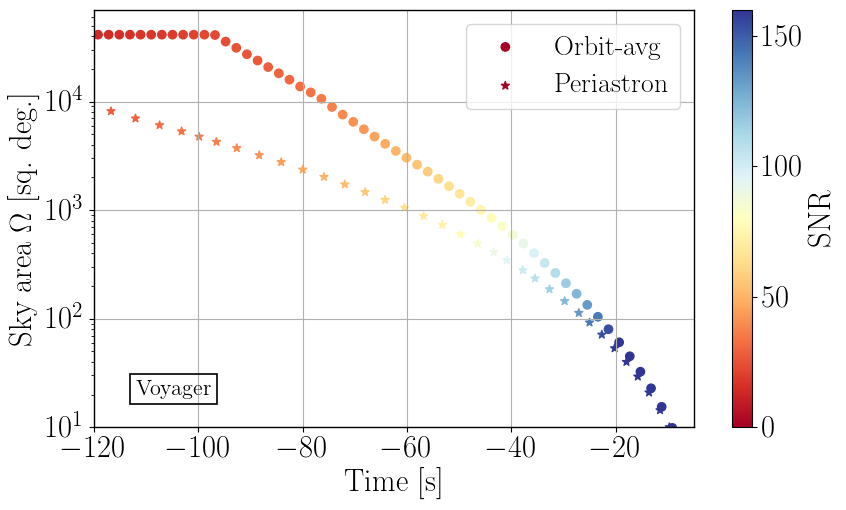}
    \end{subfigure}
    \hfill
    \begin{subfigure}[b]{0.48\textwidth}
         \centering
         \includegraphics[width=\textwidth]{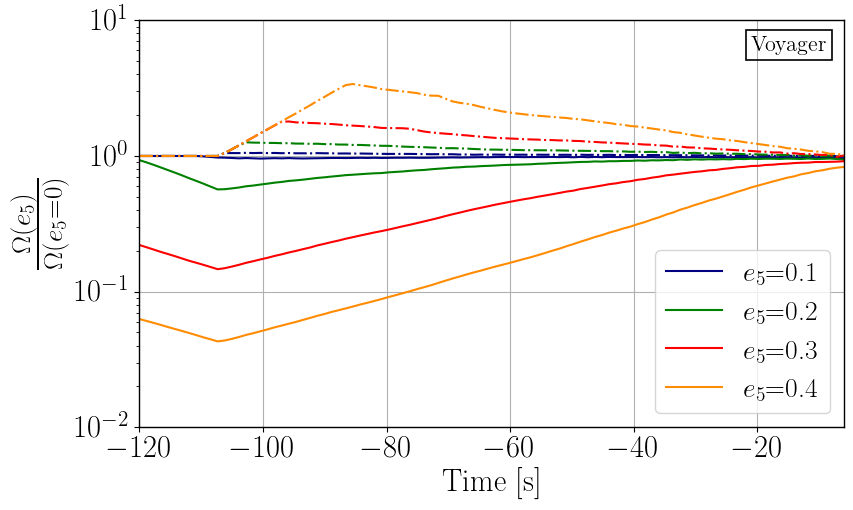}
    \end{subfigure}
    \hfill
    \begin{subfigure}[b]{0.48\textwidth}
         \centering
         \includegraphics[width=\textwidth]{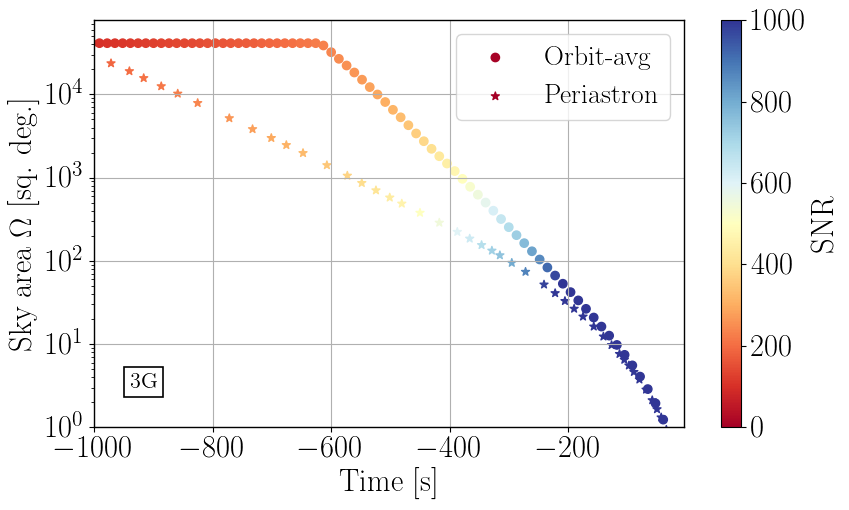}
         
    \end{subfigure}
    \hfill
     \begin{subfigure}[b]{0.48\textwidth}
         \centering
         \includegraphics[width=\textwidth]{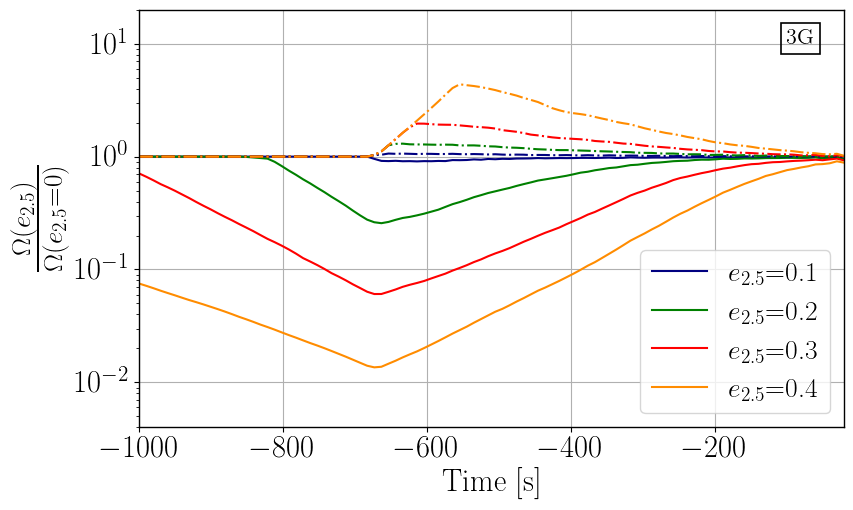}
    \end{subfigure}
    \caption{\justifying Left Column: Sky area and SNR as functions of time to coalescence for a binary with $m_1 = 10 \, M_\odot$, $m_2 = 1.4 \, M_\odot$, computed using periastron and orbit-averaged frequency evolution. Each row corresponds to a different observing scenario: O5, Voyager, and 3G, with initial eccentricities set to $e_5 = 0.3$ for O5 and Voyager, and $e_{2.5} = 0.3$ for 3G. For the SNR and sky area calculations, we consider $f_{\mathrm{start}} = 10$ Hz for O5 and Voyager and 5 Hz for 3G. At any given time, starting from periastron frequency yields smaller sky area than from orbit-averaged frequency. For a fixed sky area of 1000 sq. deg., the time gained using periastron frequency is approximately $7.8$ seconds, $13.5$ seconds, and $185$ seconds for the O5, Voyager, and 3G scenarios, respectively, with corresponding SNRs of $45$, $65$, and $381$. Right Column: Sky area ratio of eccentric to non-eccentric cases for different initial eccentricity values. Solid lines represent the sky area ratio when templates are initialized from a given periastron frequency, while the dash-dot lines represent the same using orbit-averaged frequency.}
    \label{fig:initial_condition}
\end{figure*}

We consider a fiducial NSBH binary and calculate the sky-area to which it can be localized, as a function of time to merger $t_c$. We also calculate the signal-to-noise ratio (SNR) accumulated as a function of time. We repeat these calculation using both strategies for template initial conditions. 
The first strategy (which we will call IC1 here onward) considers the signal starting time to correspond to when twice of the orbit-averaged orbital frequency enters the detector's sensitivity band, while truncating the signal at the chosen time to merger $t_c$. Likelihood is integrated between the lower and upper limits of the detector's sensitivity band. 
Referring back to the left panel inset in Fig.~\ref{fig:frequency_evolution}, this amounts to considering the waveform for integration of frequency moments for sky-localization area calculation from the vertical dot-dashed line onward, and stopping it at the time at which we want to report the sky-area and SNR.
The second approach (we will call this IC2 here onward) involves starting templates from the moment when the periastron frequency enters the detector's band and stopping it at $t_c$. This implies we start sky-area integration from the vertical solid line onward in the same inset of Fig.~\ref{fig:frequency_evolution}, and stopping again at the value of time coordinate at which we want to report the sky-area and accumulated SNR. In this case the upper frequency limit of integration is chosen to be the interpolated periastron frequency evaluated at the end time of the signal.

Our results are shown in Fig.~\ref{fig:initial_condition}. The left column illustrates the sky localization area and SNR as a function of time until coalescence for different detector configurations. In each panel
the IC1 initial condition results are shown through filled circles and that using IC2 through filled asterisks.
The detector configuration is denoted in the lower left corner of each panel.
In these panels we restrict ourselves to the dominant $\ell = |m| = 2$ mode. We consider an NSBH system with component masses $m_1 = 10\,M_\odot$ and $m_2 = 1.4\,M_\odot$, and adopt an eccentricity of $e_5 = 0.3$ for the O5 and Voyager scenarios, and $e_{2.5} = 0.3$ for 3G (additional binary parameters are detailed in Sec.~\ref{subsec: observing_scenario}). 
For O5 and Voyager, the likelihood integration begins at $f_\mathrm{start} = 10$ Hz. 

{For 3G, likehood integration starts at $5$Hz.} These details are summarized in Table~\ref{tab:table}. The SNR and sky localization area are calculated following the methodology described in Sec.~\ref{subsec:localizing_gw_sources}, fixing the probability $p=0.9$ in Eq.~\ref{eq:area} to obtain $90\%$ credible sky-areas.
Fig.~\ref{fig:initial_condition} quantitatively demonstrates that the sky area obtained using IC2 initial conditions for templates is consistently smaller than that obtained using IC1 initial conditions across all the observing scenarios.
For example, for a sky area of $10^2$ square degrees, starting search templates from periastron frequency yields a time gain of $1.4$, $2.7$ and $54$ seconds compared to the orbit-averaged frequency in the O5, Voyager and 3G scenarios respectively.
At a sky area of $10^3$ square degrees, the additional time gain is $7.8$ seconds for an SNR of $45$ (O5), $13.5$ seconds for an SNR of $65$ (Voyager), and $185$ seconds for an SNR of $381$ (3G). {\it Therefore, the neglect of these extra cycles leads to measurable deprecation of sky-localization abilities in all observing scenarios.}

 The right column of Fig.~\ref{fig:initial_condition} presents the sky area ratio of eccentric binaries compared to their quasi-circular counterparts for both initial condition strategies: orbit-averaged (dash-dotted lines) and periastron frequency (solid lines). Results are shown for initial eccentricities $e\in\{0.1, 0.2, 0.3, 0.4\}$. A sky-area ratio greater than 1 implies that the eccentric binary yields a larger localization area than the circular case, requiring follow-up telescopes to scan a broader region of the sky. When the same is less than 1 it indicates that the sky area for eccentric cases is smaller than that for non-eccentric cases, implying that the presence of eccentricity aids in localizing signals better than the circular case.
 For all the detector scenarios, {\it we can see the sky area ratio using IC1 initial conditions is always greater than 1}, and increases with increasing initial eccentricity. This is because as we increase eccentricity, the instantaneous orbit-averaged frequency decreases at any particular time, so it enters the detector band later than the non-eccentric case as shown in the inset of right panel of  Fig.~\ref{fig:frequency_evolution}. This leads to lower SNR (hence, larger sky area) for eccentric GW signals when using IC1 initial conditions for filter templates. For e.g., at time $-80\,\mathrm{s}$, the sky area of a binary with initial eccentricity $e = 0.4$ is approximately $2.9$~($3.1$) times as large as that of a circular binary in the O5 (Voyager) scenario. For the 3G scenario, the sky area is approximately $4.3$ times as large at $-554\,\mathrm{s}$. Therefore, it is evident that IC1 initial conditions is not effective for accurately localizing eccentric GW signals, compared to circular binaries.
In the same right column of Fig. \ref{fig:initial_condition}, the solid lines depict the sky area ratio of eccentric to non-eccentric signals, computed using the IC2 approach. This ratio is consistently less than 1 across all scenarios and further decreases with increasing eccentricity. This reduction occurs because highly eccentric signals oscillate at higher frequencies during each orbit, causing some cycles to enter the detector's band earlier (shown in right plot of Fig.~\ref{fig:frequency_evolution}). Additionally, while the signal duration shortens with increasing eccentricity, the amplitude increases, contributing to a higher relative SNR, as demonstrated in Ref. \cite{chen2021observation}. In summary, as eccentricity increases, the periastron frequency rises, leading to earlier band entry, and the enhanced amplitude of eccentric signals results in a higher SNR and a smaller sky area. 
As an example, in the O5 (Voyager) scenario, a NSBH binary ($m_1 = 10\,M_\odot$, $m_2 = 1.4\,M_\odot$) with an initial orbital eccentricity of 0.4 can be localized to a sky area that is approximately $1/14$ ($1/19$) that of an equivalent circular binary at $\sim -100\,\mathrm{s}$ before merger. 
Similarly, in the 3G scenario, the sky area of the eccentric binary is approximately $1/72$ that of the non-eccentric binary at $\sim -663\,\mathrm{s}$ before merger. As expected, the sky area ratio approaches 1 near merger due to the circularization of the eccentric orbit.

In all panels in the right column of Fig.~\ref{fig:initial_condition}, note that at early times, with orbit-averaged initial conditions (IC1) used to start the waveform templates, the ratio of sky area to which an eccentric binary is localized to that of a non-eccentric binary remains at unity. This is due to both configurations initially localizing the source to the {\it entire} sky ($41,253$ sq. deg.). Over time, this ratio increases from unity as the sky area for the non-eccentric case starts to diminish, while the sky area for the eccentric case remains the whole sky. Eventually, the ratio declines as the sky area for the eccentric case also begins to contract. Conversely, when starting with periastron initial conditions (IC2), the sky area for the eccentric case decreases first, while the non-eccentric case still covers the entire sky. This leads to an initial decrease in the ratio, which then reverses and increases once the sky area for the non-eccentric case {\it also} begins to decrease. This observation supports our conclusion that using the periastron frequency as the initial condition yields better sky localization for eccentric GW signals. Therefore, it should be the preferred approach for effective localization of eccentric binary GW sources by early-warning systems.

In summary, here we presented the preferred initial condition for waveform templates of GWs emitted from eccentric compact binaries. We compared two options: 1) templates starting from a given orbit-averaged frequency, and 2) templates starting from a given periastron frequency. This comparison was conducted across a range of eccentricity values and various detector configurations. Our findings indicated that using the orbit-averaged frequency as the initial condition for template generation resulted in measurably poor sky localization for eccentric binaries, especially when compared to the quasi-circular case. Conversely, initiating templates from periastron frequencies yielded significantly improved sky localization for eccentric signals relative to the quasi-circular signals.
In the remainder of this paper, we therefore use the IC2 approach to specify the initial conditions of waveforms for subsequent calculations.

\subsection{Sky localization with subdominant modes} \label{subsec: higher_mode_contribution_eccentric}

\begin{figure*}
    \centering
     \begin{subfigure}[b]{0.32\textwidth}
         \centering
         \includegraphics[width=\textwidth]{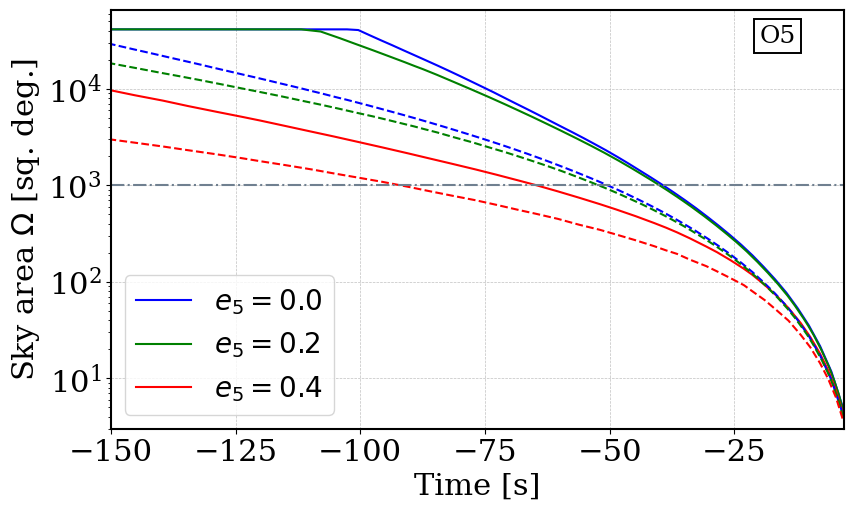}

     \end{subfigure}
     \hfill
     \begin{subfigure}[b]{0.32\textwidth}
         \centering
         \includegraphics[width=\textwidth]{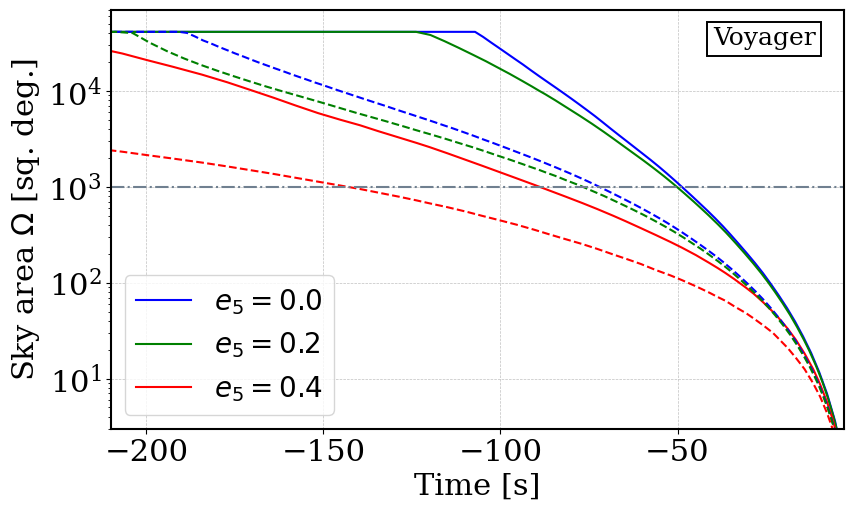}
         
    \end{subfigure}
    \hfill
     \begin{subfigure}[b]{0.32\textwidth}
         \centering
         \includegraphics[width=\textwidth]{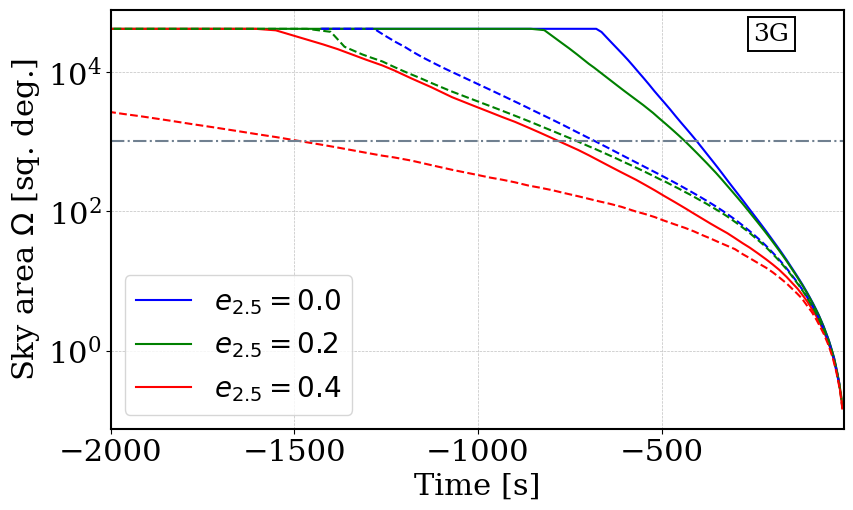}
         
    \end{subfigure}
    \hfill
    \begin{subfigure}[b]{0.32\textwidth}
         \centering
         \includegraphics[width=\textwidth]{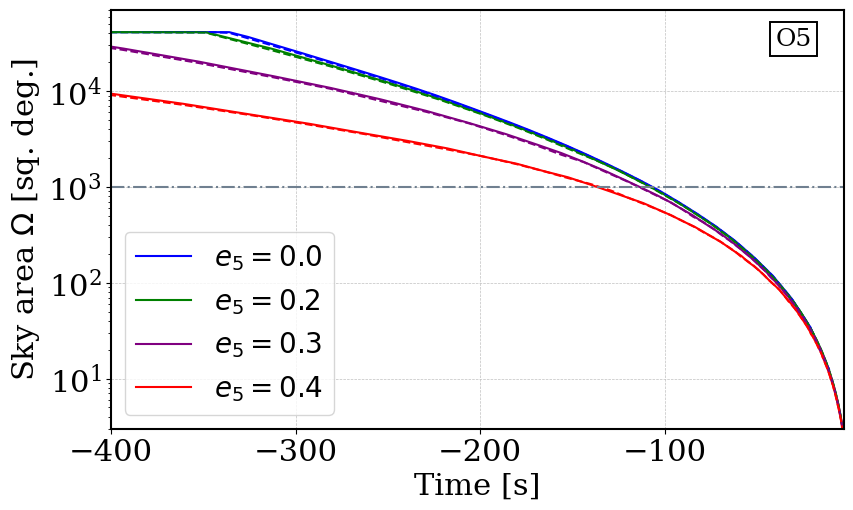}

     \end{subfigure}
     \hfill
     \begin{subfigure}[b]{0.32\textwidth}
         \centering
         \includegraphics[width=\textwidth]{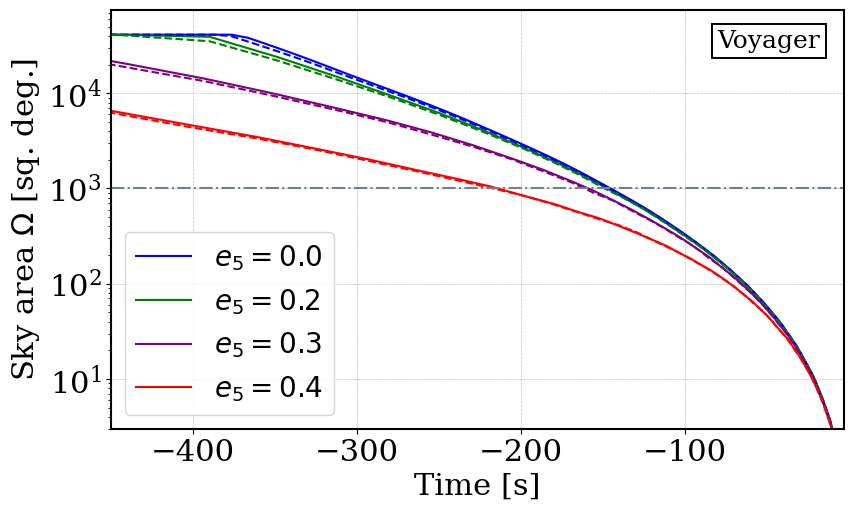}
         
    \end{subfigure}
    \hfill
     \begin{subfigure}[b]{0.32\textwidth}
         \centering
         \includegraphics[width=\textwidth]{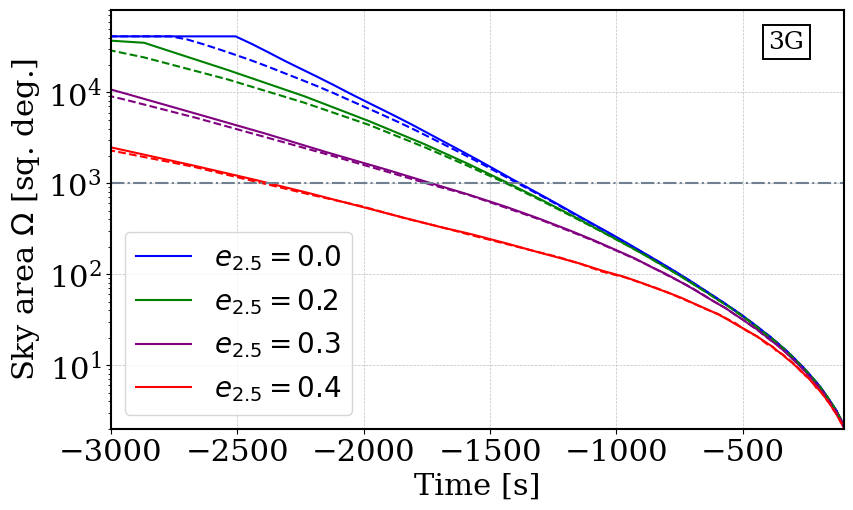}
         
    \end{subfigure}
    \caption{\justifying Sky localization area as a function of time to coalescence for different initial eccentricities, shown for various observing scenarios. Top Panel: NSBH binary system with $m_1= 10 \, M_\odot$, $m_2= 1.4 \, M_\odot$, and $\chi_1=\chi_2=0$. Solid lines represent the sky area calculated using only the dominant $(2,2)$ mode, while the dashed lines include both dominant $(2, 2)$ and subdominant modes. For the $e=0$ case, at 1000 sq. deg sky area, time gain using subdominant alongside dominant mode is $11$ seconds, $23$ seconds, and $4.6$ minutes for O5, Voyager, and 3G scenarios. For the $e=0.4$ case, this early warning time gain increases to $27$ seconds, $54$ seconds, and $11.6$ minutes, respectively. Bottom Panel: BNS system with $m_1= m_2= 1.4 \,M_\odot$. While subdominant modes offers only marginal improvements in the sky area for these binaries, orbital eccentricity notably tightens sky localization compared to quasi-circular systems.
    }
    \label{fig:area 22 and HM at different eccentricity}
  \end{figure*}

Section \ref{subsec: initial_conditions_for_eccentric_templates} demonstrated the advantage of using the periastron frequency to specify template initial conditions and improving the sky localization of eccentric GW sources. Based on that, this section onward we proceed with sky localization calculations that utilize the periastron frequency as the initial condition (i.e. strategy IC2).

In this section we will quantify the significance of subdominant modes relative to the dominant GW modes from eccentric binary mergers when it comes to early-warning. From Fig.~\ref{fig:hm_contribution_eccentric_gws}, we note that the $(\ell, |m|) =(3,3), (4,4)$, and $(2,1)$ modes contribute at least $2\%$ of the total SNR and other modes contribute lesser. Therefore, in our subsequent calculations, we will include these three subdominant modes alongside the dominant $(2,2)$ mode. The specific initial parameters for different detector scenarios are listed in Table \ref{tab:table}.

We start with the same fiducial NSBH binary as we considered in the previous subsection. Top panel of Fig. \ref{fig:area 22 and HM at different eccentricity} illustrates the sky localization with and without subdominant modes for various initial eccentricities for that binary with masses $m_1=10 \, M_\odot, m_2=1.4 \, M_\odot$. As before we consider O5, Voyager, and 3G scenarios (detailed in Sec.~\ref{subsec: observing_scenario}). Solid lines represent sky localization using only the dominant mode, while dashed lines represent sky localization using all considered modes. The different colors represent different initial eccentricity values.
Similar to the circular case, where the inclusion of subdominant modes reduces the sky area (as demonstrated in \cite{kapadia2020harbingers, singh2021improved, singh2022improved}), we observe that incorporating subdominant modes with the dominant mode also decreases the sky area for eccentric GW signals. Furthermore, increasing initial eccentricity leads to an even greater reduction in the sky area relative to the circular case, both with and without the inclusion of subdominant modes. 
We found that the early warning time gain for $e=0.4$ binary systems with respect to the $e=0$ case, is approximately $25$ seconds, $40$ seconds and $6.2$ minutes for O5, Voyager and 3G scenarios, respectively at a sky area of 1000 square degrees when considering only the dominant $(2,2)$ mode.  When sub-dominant modes are included alongside the dominant mode, this time gain for $e=0.4$ binaries, relative to $e=0$, increases to $41$ seconds, $71$ seconds and $13$ minutes for the respective observing scenarios.
Furthermore, for an eccentric system with $e=0.4$, the early warning time gain achieved specifically by including sub-dominant modes, compared to their exclusion, is $27$ seconds, $54$ seconds, and $11.6$ minutes at a fiducial sky area of 1000 square degrees for the O5, Voyager, and 3G scenarios, respectively.

Bottom panel of Fig.~\ref{fig:area 22 and HM at different eccentricity} illustrates the same but for BNS systems with $m_1= m_2=1.4 \, M_\odot$. For BNS, as expected, the effect of including subdominant harmonics is less prominent. The effect of orbital eccentricity is manifestly more prominent, and overtakes the effect of including subdominant harmonics at higher eccentricities. Eccentric BNS will therefore get substantially better localized in low latency than their quasi-circular counterparts, leading to longer early-warning times.

\begin{figure*}
    \centering
    \includegraphics[width=0.9\linewidth]{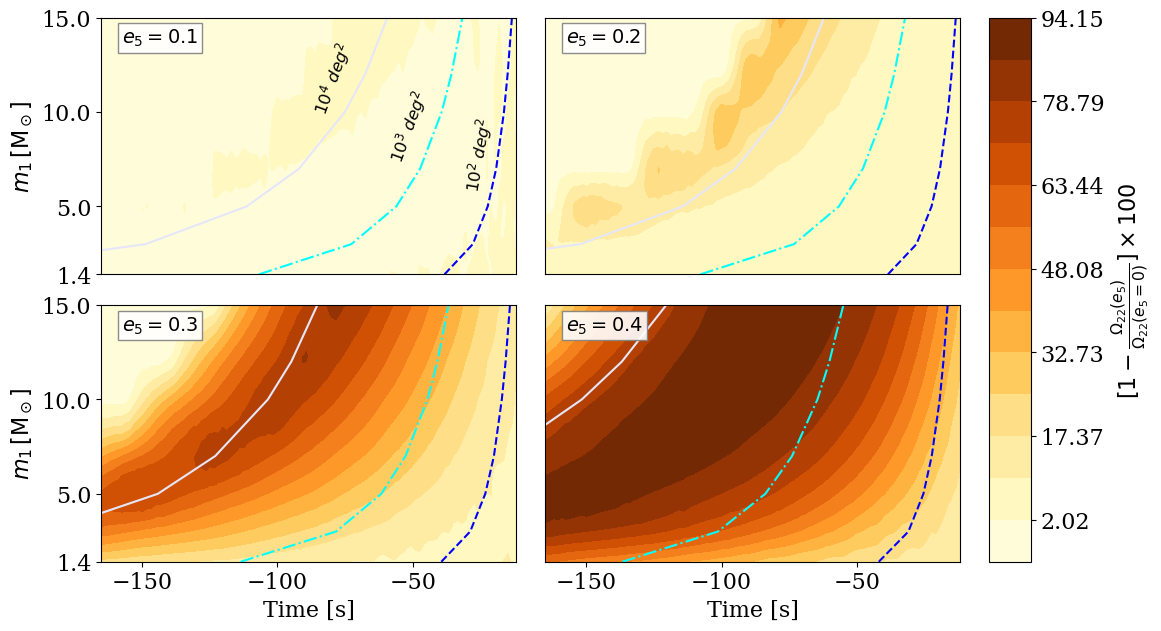}
    \caption{\justifying The fractional reduction of sky area due to eccentricity, given by Eq.~\eqref{eq:fractional_reduction}, is shown using only the dominant $(2,2)$ mode for O5 Scenario. The contours correspond to the absolute sky area, with white, cyan, and blue lines indicating absolute sky areas for eccentric signals of $10^4, 10^3$, and $10^2$ sq. deg., respectively. 
    }
    \label{fig:ratio_eccentric_to_circ_22}
\end{figure*}

We next extend our investigation by exploring the effect of subdominant modes on the sky localization of eccentric GW signals from binary neutron star and neutron star-black hole binaries for a variety of binary component masses and spins.
Fig.~\ref{fig:ratio_eccentric_to_circ_22} shows the fractional reduction in sky area when considering the dominant mode for eccentric binaries compared to circular binaries for the O5 scenario. Different panels correspond to different initial orbital eccentricities, as shown in their top left corner. The x-axis of all panels represents the time to coalescence, and the y-axis represents the primary mass $m_1 \in[1.4,15]$, with the secondary mass $m_2=1.4 \, M_\odot$ and zero spins on both compact objects. For all our calculations here, we fixed the secondary object's mass to $1.4 \, M_\odot$. The color bar indicates the fractional reduction in sky area $\Delta\Omega_e$, with
\begin{align}\label{eq:fractional_reduction}
    \Delta\Omega_e(\%) = \frac{\Omega(e=0) - \Omega(e)}{\Omega(e=0)}\times 100.
\end{align}
For eccentricities below 0.4, $\Delta\Omega_e$ is initially close to zero during the early inspiral phase. This reduction increases to a maximum at some point during the inspiral and then decreases. The small reduction at the beginning of the inspiral occurs because the sky area is at its maximum ($41,253$ square degrees) for both eccentric and non-eccentric binaries. This means both eccentric and non-eccentric binaries have not yet been meaningfully localized on the sky.
$\Delta\Omega_e$ reaches a maximum as the inspiral progresses. Subsequently, it decreases as the GW signals approach merger, eventually returning to smaller values closer to merger. This is because by the end both have nearly accumulated all their SNR\footnote{In the O5 scenario, and considering only the dominant mode, the total SNR accumulated by an eccentric NSBH binary (with $m_1=10 \, M_\odot, m_2=1.4 \, M_\odot$) with an eccentricity of $0.4$ is $167$, whereas for a quasi-circular binary, this becomes $160$.  
}, and so the final fractional reduction is simply a property of the relative SNRs of eccentric and non-eccentric sources. 
The contour lines on the plot represent the {\it actual sky-localization areas for the eccentric signals}, corresponding to $10^4, 10^3$, and $10^2$ square degrees. They help indicate the practically useful regions of the figure. For e.g., actual sky-localization areas larger than $1,000$ square degrees might be considered too extensive for effective signal localization.
We observe that as the initial eccentricity ($e_5$) increases from $0.1$ to $0.4$ for NSBH systems, the fractional reduction in sky area, at an actual 
sky-localization area (corresponding to eccentric signals) of $1,000$ square degrees (dash-dotted contours), increases from less than $2\%$ to $80\%$. For the fiducial NSBH binary with $m_1=10 \, M_\odot$ and $m_2=1.4 \, M_\odot$ we find an extra early warning time gain of $25$ seconds.
For an absolute sky-localization area of $100$ square degrees, this reduction ranges from $1\%$ to $30\%$, with an approximate early warning time gain of $4$ seconds. For the BNS system (bottom edge of each panel), the fractional improvement in sky-localization area ranges from $1\%$ to $40\%$ when the source has been localized to within $1,000$ sq. deg. with initial eccentricities going from $e_5=0.1$ to $e_5=0.4$ with an extra early warning time of 30 seconds. Furthermore, increasing the eccentricity expands the region in parameter space where the sky area reduction is high. 

\begin{figure*}
    \centering
    \begin{subfigure}[b]{0.75\textwidth}
         \centering
         \includegraphics[width=\textwidth]{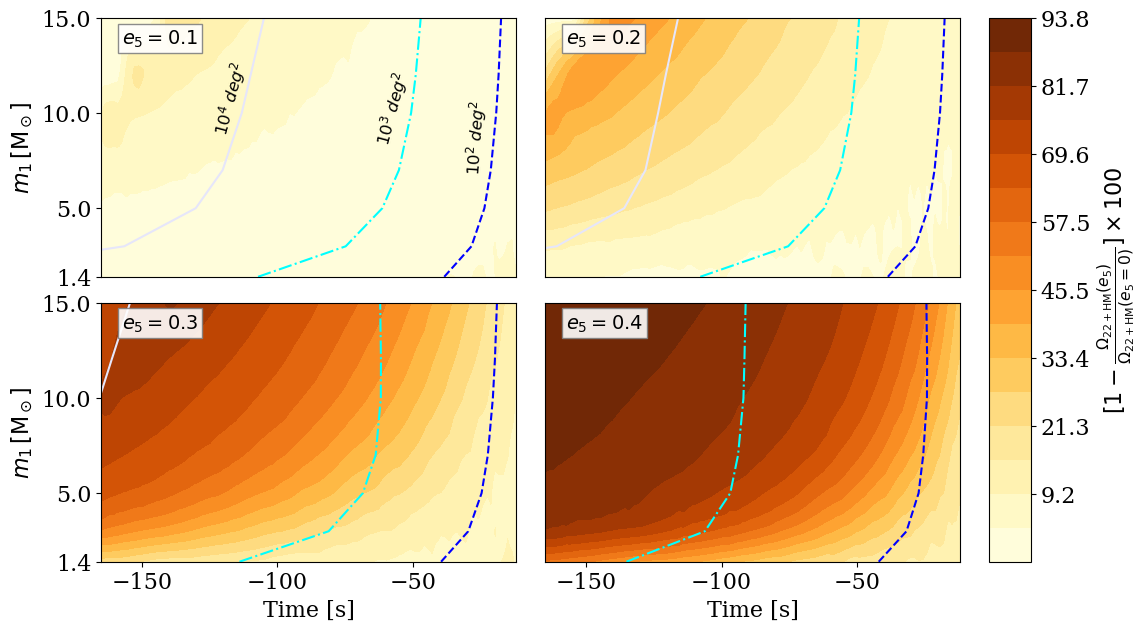}
         \caption{\textbf{O5}}

     \end{subfigure}
     \hfill
     \begin{subfigure}[b]{0.75\textwidth}
         \centering
         \includegraphics[width=\textwidth]{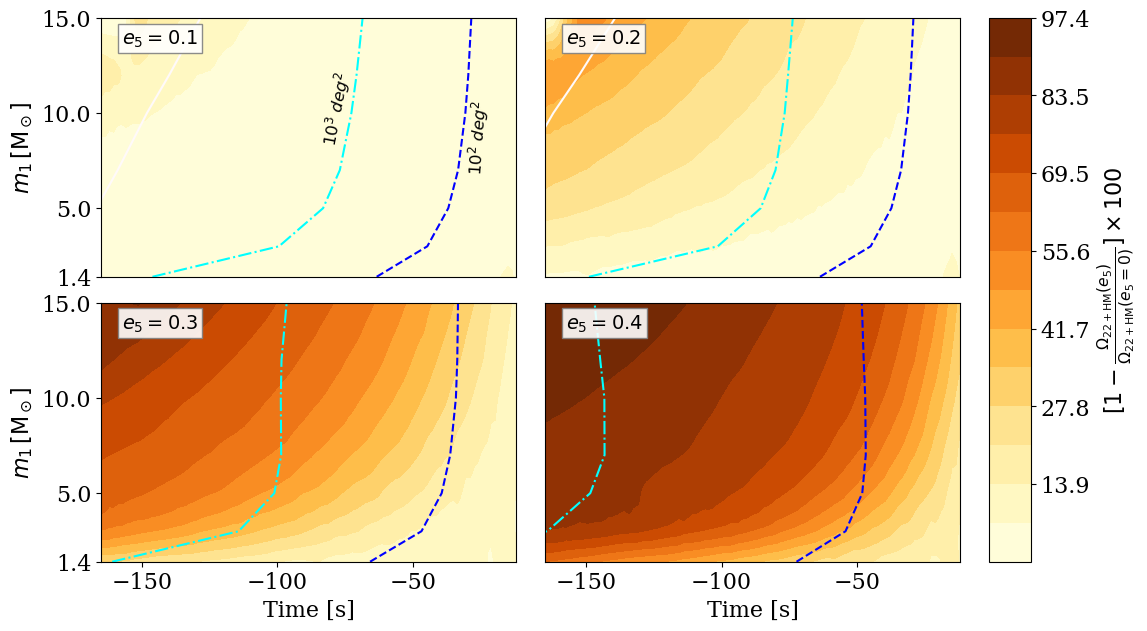}
         \caption{\textbf{Voyager}}
         
    \end{subfigure}
   
    \caption{\justifying Fractional reduction of sky area due to eccentricity, computed using subdominant modes along with dominant mode, is shown for different initial eccentricities in the O5 and Voyager scenarios. The horizontal axis represents the time to coalescence, while the vertical axis is the mass of the primary compact object in binary. The white, cyan, and blue contours represent absolute sky areas of $10^4$, $10^3$, and $10^2$ sq. deg., respectively. 
    }
    \label{fig:ratio_eccentric_to_circ_22+HM}
\end{figure*}

Fig.~\ref{fig:ratio_eccentric_to_circ_22+HM} presents the ratio of sky areas for eccentric versus non-eccentric binaries, extending the analysis of Fig.~\ref{fig:ratio_eccentric_to_circ_22} by considering {\it both the dominant and subdominant GW modes} for O5 and Voyager scenarios. As before, this figure includes both BNS and NSBH systems. The contour lines represent the {\it actual sky-localization areas for the eccentric binaries}, corresponding to $10^4, 10^3$, and $10^2$ square degrees. 
In the O5 scenario, the fractional reduction in sky area remains comparable to that shown in Fig.~\ref{fig:ratio_eccentric_to_circ_22}. As eccentricity increases from \( e_5 = 0.1 \) to \( e_5 = 0.4 \), we observe a gain in early warning time of approximately $6$ seconds at $100$ sq. deg. for a fiducial NSBH system of $(10, 1.4) \, M_\odot$. For the Voyager scenario, we get somewhat similar reductions due to eccentricity but overall better sky localization since all contours move to earlier times.
At $100$ sq. deg., we get around $2\%$ fractional reduction in sky area for $e_5=0.1$ at $t_c= -31$ s, whereas for $e_5=0.4$ eccentricity, this reduction goes up to $65\%$ at $t_c=-47$ s for the same NSBH system. So the extra early warning time gain from $e_5=0.1$ to $e_5=0.4$ is around $16$ seconds at $100$ sq. deg. For the BNS system, this additional time gain with increasing eccentricity is around $10$ seconds.

\begin{figure*}
    \centering
     \begin{subfigure}[b]{0.75\textwidth}
         \centering
         \includegraphics[width=\textwidth]{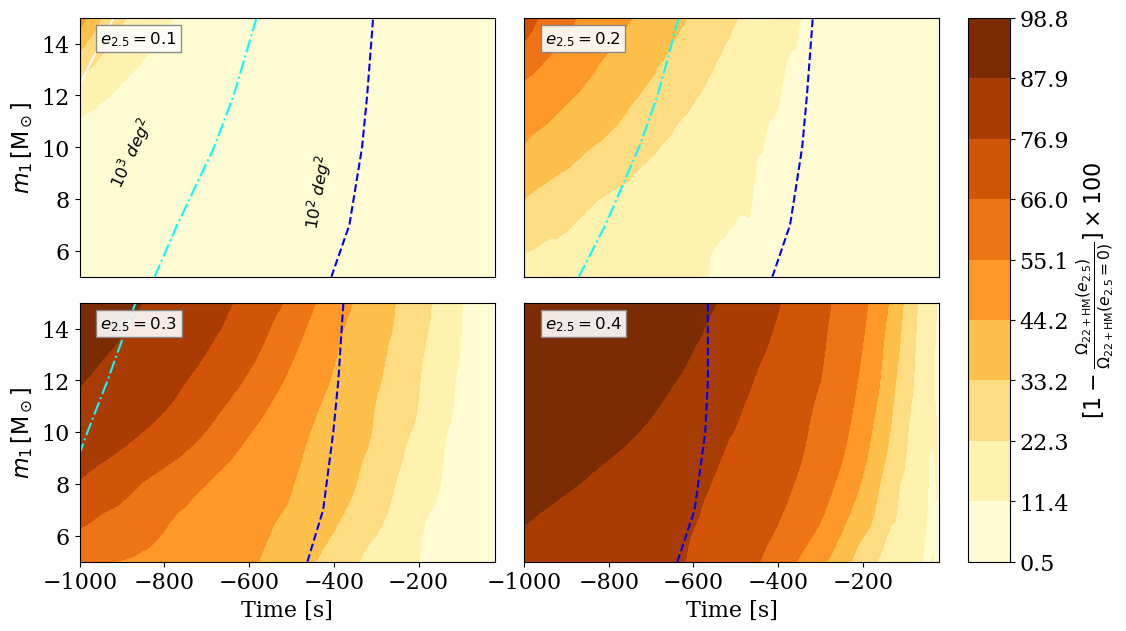}
         \caption{\textbf{3G: NSBH}}
    \end{subfigure}
    \hfill
    \begin{subfigure}[b]{0.75\textwidth}
         \centering
         \includegraphics[width=\textwidth]{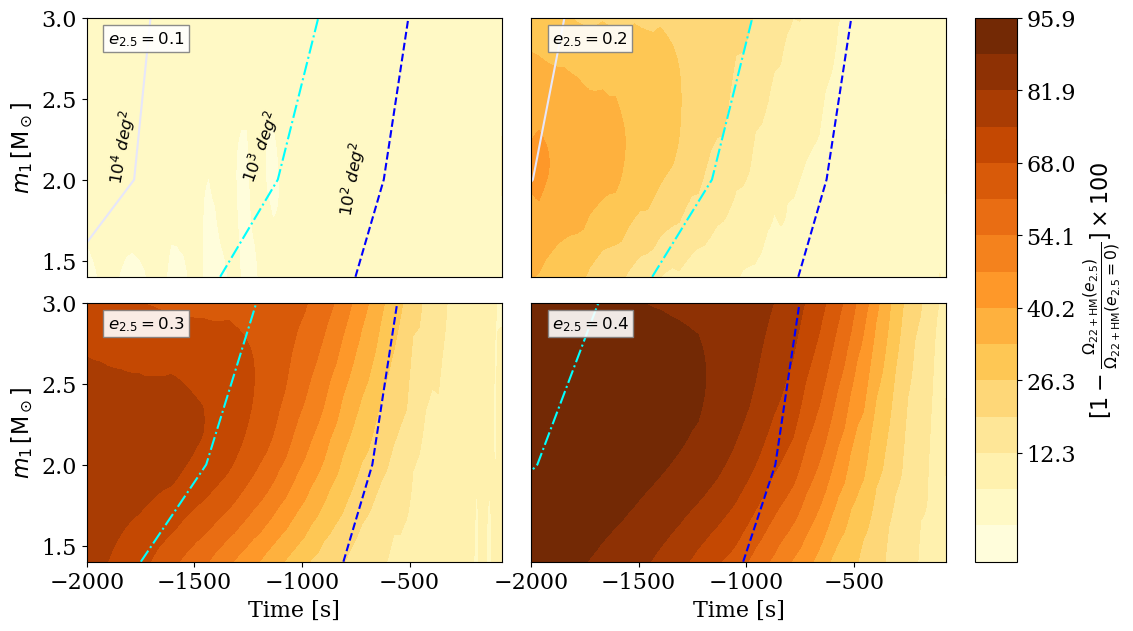}
         \caption{\textbf{3G: BNS}}
    \end{subfigure}
    
    \caption{\justifying Same as Fig.~\ref{fig:ratio_eccentric_to_circ_22+HM}, but for the 3G scenario. The sky area results for BNS and NSBH binaries are presented separately. For NSBH systems, we use the complete inspiral-merger-ringdown waveform, whereas for BNS systems, we use only the inspiral portion of the waveform. 
    }
    \label{fig:ratio_eccentric_to_circ_22+HM_3G}
\end{figure*}

Fig.~\ref{fig:ratio_eccentric_to_circ_22+HM_3G} presents the same analysis as Fig.~\ref{fig:ratio_eccentric_to_circ_22+HM} but for the 3G scenario. Here, we present the results for NSBH and BNS systems separately. As before, complete inspiral-merger-ringdown waveforms are used for the NSBH systems, whereas for BNS systems, only the inspiral portion is considered due to the high computing memory requirements of long-duration waveforms starting at low frequencies. The waveforms are generated with an initial eccentricity defined at $2.5$ Hz, and the $f_{start}$ for SNR integration is $5$ Hz, detailed in Table.~\ref{tab:table}. For NSBH ($m_1=10 \, M_\odot, m_2=1.4 \, M_\odot$), at sky area of $100$ sq. deg., the fractional reduction for eccentric ones with $e_{2.5}=0.1$ is about $2\%$ relative to circular binaries at $t_c=-334$ s, and for $e_{2.5}=0.4$, this reduction is $80\%$ at $t_c=-572$ s. The extra time gain at $100$ sq. deg., because of increasing eccentricity (from $0.1$ to $0.4$) is $238$ seconds ($\sim 3.97$ minutes).
For BNS systems (with $m_1=m_2=1.4 \, M_\odot$), we can get {\it $40$ minutes} of early warning time for $e_{2.5}=0.4$ eccentricity at $1000$ sq. deg sky area with fractional reduction of $90\%$, while at $100$ sq. deg. sky area, this early warning time is {\it $16.8$ minutes} with fractional improvement in sky-localization of eccentric case with $e_{2.5}=0.4$ being about $60\%$.
From these results we conclude that, though possibly rare, eccentric binaries with neutron stars (NSBH/BNS) can have better early sky-localization and therefore have better chances of being followed up for prompt electromagnetic emissions than their quasicircular counterparts. Based on this we recommend that efforts within the early-warning GW community consider low-latency early-warning efforts targeted at these eccentric binaries.

\begin{figure*}
    \centering
    \begin{subfigure}[b]{0.80\textwidth}
         \centering
         \includegraphics[width=\textwidth]{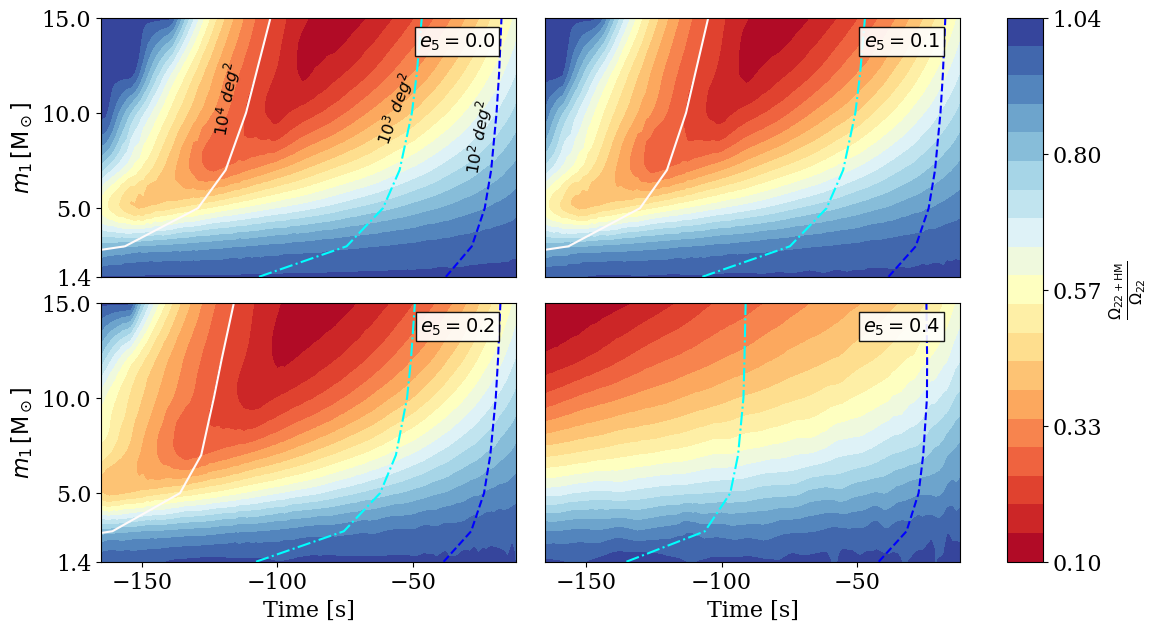}
         \caption{\textbf{O5}}

     \end{subfigure}
     \hfill
     \begin{subfigure}[b]{0.80\textwidth}
         \centering
         \includegraphics[width=\textwidth]{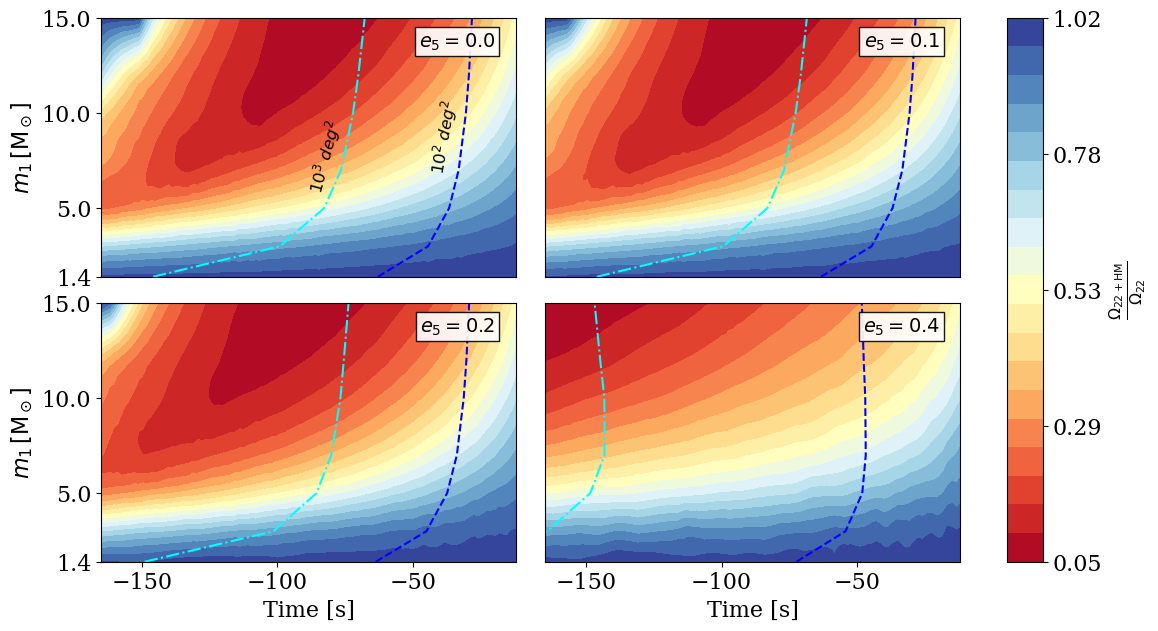}
         \caption{\textbf{Voyager}}
         
    \end{subfigure}
    
    \caption{\justifying $\frac{\Omega_{(22+HM, e)}}{\Omega_{(22, e)}}$ for a range of primary BH's mass ($m_1$) while keeping $m_2=1.4 \, M_\odot$ at different eccentricity value, for O5 and Voyager scenario. 
    }
    \label{fig:ratio 22+hm to 22 mode O5}
\end{figure*}

\begin{figure*}
    \centering
    
     \begin{subfigure}[b]{0.80\textwidth}
         \centering
         \includegraphics[width=\textwidth]{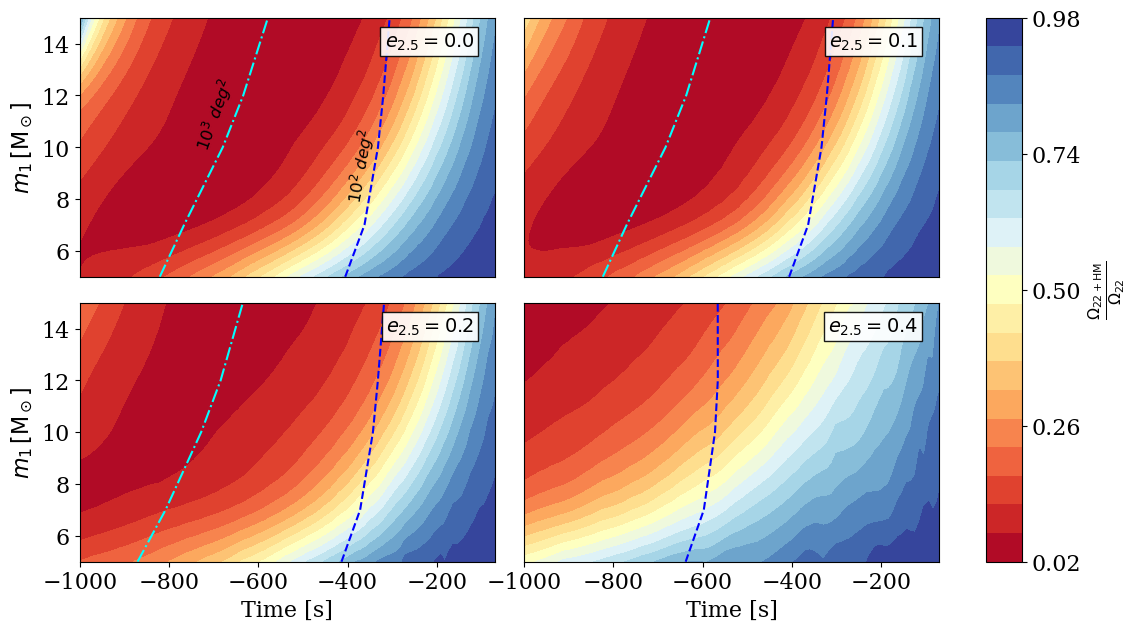}
         \caption{\textbf{3G: NSBH}}
         
    \end{subfigure}
    \hfill
    \begin{subfigure}[b]{0.80\textwidth}
         \centering
         \includegraphics[width=\textwidth]{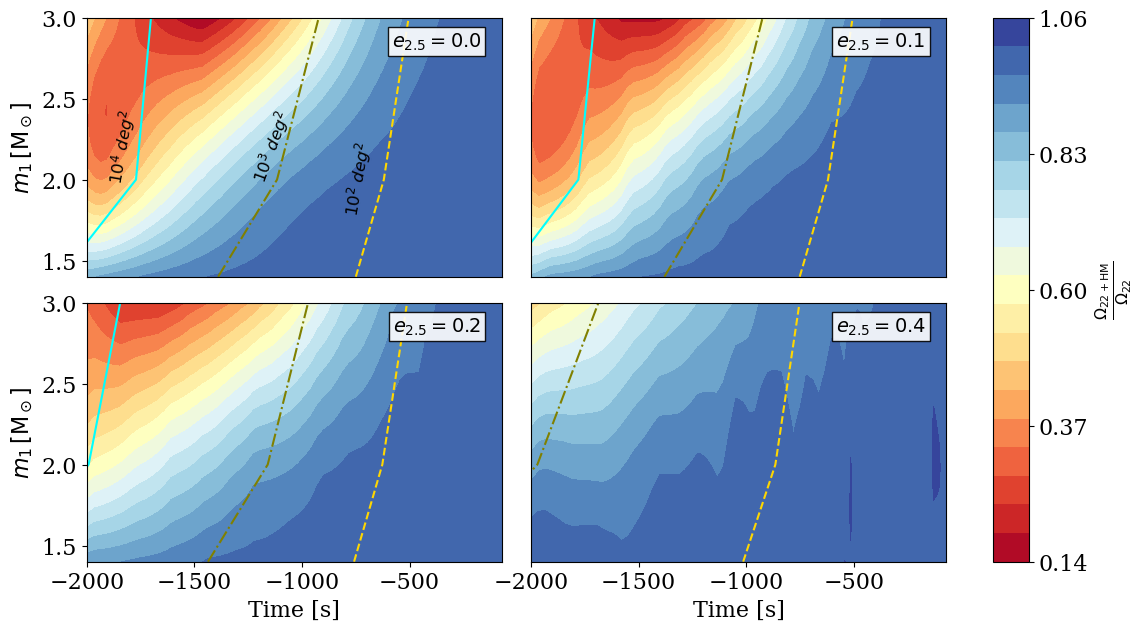}
         \caption{\textbf{3G: BNS}}
         
    \end{subfigure}
    \caption{\justifying $\frac{\Omega_{(22+HM, e)}}{\Omega_{(22, e)}}$ for a range of primary BH's mass ($m_1$) while keeping $m_2=1.4 \, M_\odot$ at different eccentricity value, for 3G scenario. (a) is for NSBH bianry system while (b) is for BNS system.}
    \label{fig:ratio 22+hm to 22 mode 3G}
\end{figure*}

\begin{figure*}
    \centering
    \includegraphics[width=1.0\linewidth]{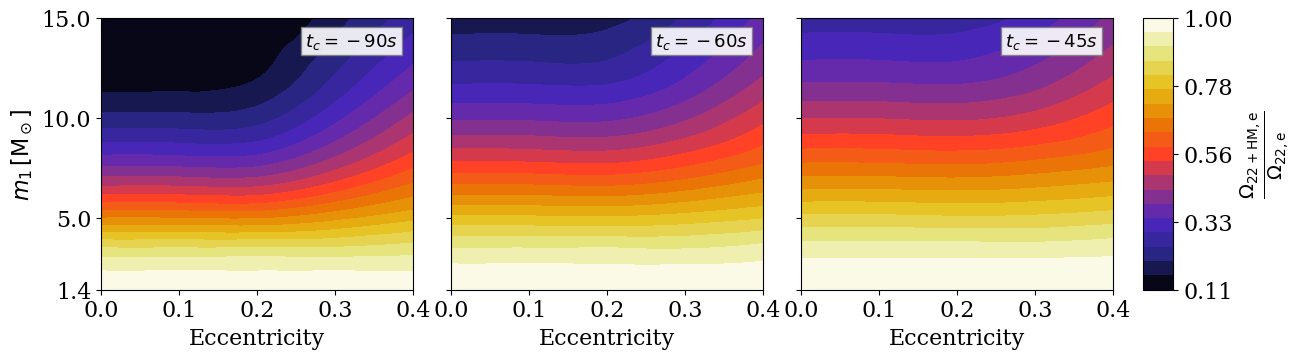}
    \caption{\justifying Row shows $\Omega_{22+HM,e}/\Omega_{22,e}$
     the coloums are for three different $t_c$ value=[-90, -60, -45] sec for O5 scenario.}
    \label{fig:HM_impacts_at_fixed_tc}
\end{figure*}

\begin{figure*}
    \centering
    \includegraphics[width=1.0\linewidth]{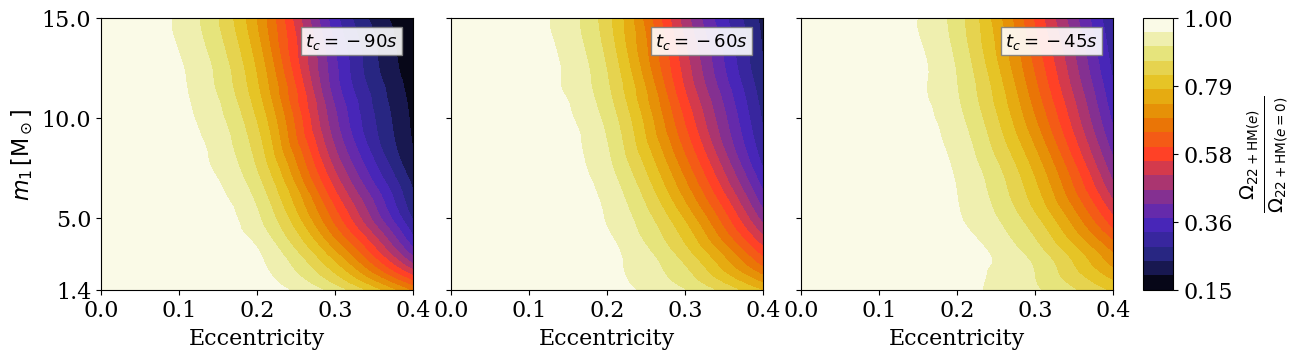}
    \caption{\justifying Row shows $\Omega_{22+HM,e}/\Omega_{22+HM,e=0}$ and the coloums are for three different $t_c$ value=[-90, -60, -45] sec for O5 scenario.}
    \label{fig:eccentricity_impacts_at_fixed_tc}
\end{figure*}

\begin{figure*}
    \centering
    \includegraphics[width=1.0\linewidth]{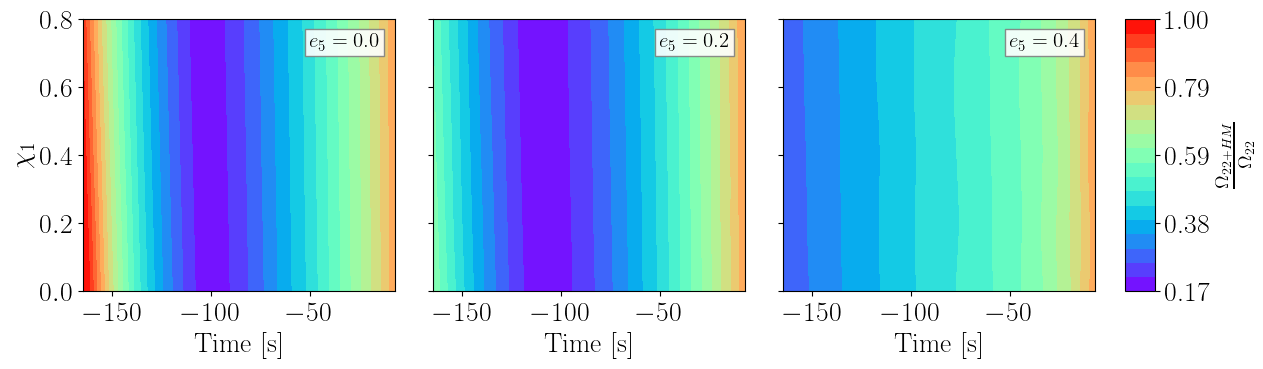}
    \caption{\justifying Sky area ratio of 22+HM to 22 mode as a function of primary spin and time to merger, for different eccentricity values, for an NSBH binary with $m_1= 10 \, M_\odot$ and $m_2= 1.4 \, M_\odot$. The eccentricities are defined at $f_{low}= 5 \, Hz$. Here we consider the O5 scenario.
    }
    \label{fig:sub-dominant_on_spin}
\end{figure*}

\begin{figure}[h]
    \centering
    \includegraphics[width=8.8cm]{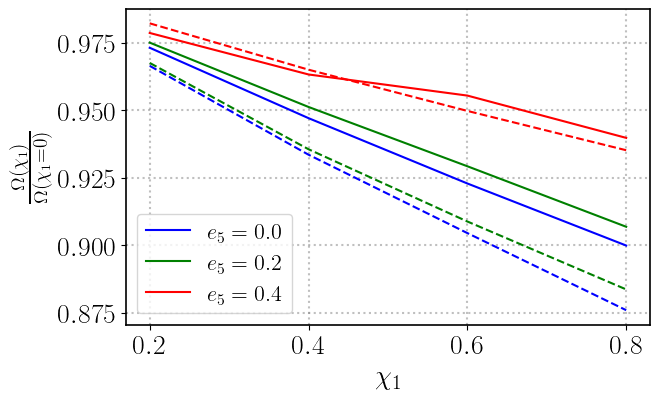}
    \caption{Sky area ratio of spinning to non-spinning for 22 and 22+HM, for three different initial eccentricity at 5 Hz, the colors shows different eccentricity value at $t_c= -45 s$, the solid line for the 22+HM and the dashed line for 22 mode, for O5 scenario.}
    \label{fig:spinning_to_non_spinning}
\end{figure}

Fig.~\ref{fig:ratio 22+hm to 22 mode O5} illustrates the impact of subdominant modes relative to the dominant mode across the parameter space of the primary mass at various eccentricities for the O5 and Voyager scenarios. The color bar indicates the ratio of sky-localization areas with versus without subdominant modes. This is distinct from the previous figures which quantified the effect of orbital eccentricity instead. Here, a smaller ratio indicates a stronger influence of subdominant modes, leading to better sky localization. In the early inspiral, the ratio is close to one as both sky areas are their maximum (i.e. no meaningful localization in either case).
Across scenarios, the sky area ratio decreases with increasing $m_1$ (while $m_2= 1.4 \, M_\odot$), indicating that, as we expect, the contribution of subdominant modes becomes more significant with increasing mass ratio. At a mass ratio of $10$, including subdominant modes can lead to approximately a $70\%(90\%)$ reduction in sky area about $48(69)$ s before merger at 1000 square degrees sky area with initial eccentricity, $e_5=0.1$, in the O5(Voyager) scenario. As eccentricity increases, the region of maximum gain from subdominant modes shifts to earlier times because higher eccentricity causes the GW signals to oscillate at higher frequencies and enter the detector band sooner. For the same binary with mass ratio 10 with $e_5=0.4$, the reduction in sky area remains close, observed at earlier times of $91(145)$s before merger for O5(Voyager) scenario.
For the BNS systems, the impact of subdominant modes is expectedly minimal, with the sky area ratio staying close to unity (consistent with the trends seen in Fig.~\ref{fig:hm_contribution_eccentric_gws}). 
Fig.~\ref{fig:ratio 22+hm to 22 mode 3G} extends this analysis to the 3G scenario and presents results for NSBH and BNS systems separately. For NSBHs, {\it we can get up to $98\%$ reduction in sky area by including subdominant modes to localize sources with mass ratio $10$}. Similar to Fig.~\ref{fig:ratio 22+hm to 22 mode O5}, as we increase eccentricity the effectiveness of subdominant modes shifted to early inspiral times. For most BNS system with mass ratio near unity, we see the reduction due to subdominant mode remains less than $3\%$, whereas for more asymmetric neutron stars binaries with $m_1>2 \, M_\odot$, a reduction of up to $20\%$ is achievable for an eccentricity of $e_{2.5}=0.4$ at early-warning times of nearly half an hour.

The impact of the primary mass ($m_1$) and eccentricity on sky localization is explored in a different way in Figs.~\ref{fig:HM_impacts_at_fixed_tc} and \ref{fig:eccentricity_impacts_at_fixed_tc}, respectively, by looking at three distinct moments until coalescence: $t_c= -90$s, $-60$ s, and $-45$ s. Fig.~\ref{fig:HM_impacts_at_fixed_tc} highlights the contribution of subdominant GW modes, whereas Fig.~\ref{fig:eccentricity_impacts_at_fixed_tc} depicts the influence of eccentricity across the relevant parameter space. As seen in Fig.~\ref{fig:HM_impacts_at_fixed_tc}, the impact of subdominant modes on sky localization becomes more pronounced with a larger primary BH mass ($m_1$). Moreover, this enhancement is even greater at earlier inspiral stages ($t_c=-90$ s) as compared to the later inspiral stages ($t_c=-45$ s), and is simultaneously lower for higher eccentricities. These are expected trends, as higher-modes become louder with increasing mass ratio leading to better sky localization, and also the same improvement can be achieved by turning up the initial orbital eccentricity. 
Quantitatively, the maximum reduction in sky localization area achieved by including subdominant modes is $89\%$ at $t_c=-90$ s, whereas it is $73\%$ at $t_c=-45$ s. 
In Fig.~\ref{fig:eccentricity_impacts_at_fixed_tc}, we investigate the impact of eccentricity by showing the ratio of sky area for eccentric to non-eccentric binaries using subdominant modes along with the dominant mode for three different times to coalescence. The vertical and horizontal axes represent the primary mass $m_1$ and eccentricity, respectively. 
With increasing eccentricity, the sky area ratio of eccentric to non-eccentric cases decreases, highlighting the impact of eccentricity. For an eccentricity of $0.4$, the sky area reduction relative to the non-eccentric case reaches approximately $84 \%, 77 \%$ and $ 71\%$  at a time to coalescence value of $t_c= -90$ s, $-60$ s, and $-45 $ s. As in Fig.~\ref{fig:HM_impacts_at_fixed_tc} this enhancement in sky localization due to eccentricity is greater at earlier inspiral stages ($t_c=-90$ s) as compared to the later inspiral stages ($t_c=-45$ s). Additionally, for a fixed eccentricity, increasing the primary mass yields only a modest improvement in the eccentricity-induced enhancement of the sky localization.
From these two figures, we conclude that {\it both mass-ratio and orbital eccentricity comparably enhance the importance of subdominant waveform harmonics, with perhaps the eccentricity playing a more important role at lower mass ratios, and consequently lead to better sky localization of their sources}.This is one of the main results of this paper.

So far, our discussion has focused on the impact of eccentricity and the subdominant mode on sky localization within the mass parameter space. Lastly, we examine these effects across a range of the heavier black hole's spin ($\chi_1$), assuming no spin on the neutron star. We consider the dimensionless primary spin $\chi_1 \in [0, 0.8]$ at various eccentricity values $e_5 = \{0, 0.2, 0.4 \}$ for a fiducial NSBH binary system with masses $m_1 = 10 \, M_\odot$, $m_2 = 1.4 \, M_\odot$, and $\chi_2 = 0$ in the O5 scenario\footnote{We defer this question for 3G scenario due to computational constraints, but we expect the overall trend to be similar, i.e. the effect of spins is subdominant to that of orbital eccentricity and of subdominant harmonics in the GW signal.}. Fig.~\ref{fig:sub-dominant_on_spin} illustrates the influence of the subdominant mode as a function of the heavier black hole's spin $\chi_1$. Consistent with our observations in Fig.~\ref{fig:hm_contribution_eccentric_gws}, where the SNR of the subdominant mode exhibited a very weak dependence on black hole spin, we see a similar trend here. With increasing spin, the impact of subdominant modes on the sky area remains nearly constant for a given $t_c$ across all considered eccentricity values. However, as eccentricity increases, the region where this effect is most pronounced (maximum reduction in sky area) shifts towards earlier times. In Fig.~\ref{fig:spinning_to_non_spinning}, we demonstrate the improvement in sky area achieved solely by including spin effects, that is, we present $ \frac{\Omega_{\chi_{1}}}{\Omega_{\chi_1=0}}$. Solid and dashed lines show results with and without subdominant modes. Notably, for zero eccentricity, the reduction in sky area is maximal for the dominant mode, showing approximately a $12\%$ reduction. The inclusion of the subdominant mode provides a reduction of around $10\%$. This reduction diminishes as we transition to higher eccentricities; for $e_5 = 0.4$, the sky area reduction due to spin is approximately $6\%$ with and without subdominant mode. Therefore, we can conclude that the black hole's spin has a relatively smaller influence on sky localization and early warning time.

\section{Conclusions}\label{sec:conclusion}
We investigated early warning signals from eccentric NSBH and BNS binary systems, given their significant potential as sources for multi-messenger astronomy. Focusing on eccentric systems is crucial, as they are expected to account for a considerable fraction of detectable mergers in upcoming observational runs and during the era of third-generation (3G) detectors. 

A key question we addressed concerns the optimal initial starting point for generating eccentric waveform templates for SNR integration. We considered two methods: one using the orbit-averaged frequency as the initial condition and the other using the periastron frequency for waveform generation. Our results indicate that initiating waveform generation and SNR integration with the orbit-averaged frequency leads to a larger sky area and a lower SNR, and this effect becomes more significant with increasing eccentricity. In contrast, using the periastron frequency as the initial condition results in a tighter sky localization compared to the orbit-averaged frequency method. Moreover, the sky localization even improves as the eccentricity increases. We found that we can gain an extra early warning time of 7.8 to 185 seconds for a binary with $m_1=10 \, M_\odot, m_2=1.4 \, M_\odot$ with an initial eccentricity of $0.3$, moving from the O5 to the 3G detector scenario at a 1000 sq. deg. sky area, simply by using the periastron frequencies as the initial condition instead of the orbit-averaged frequency.

This highlights an essential point for accurately locating eccentric GW sources in the sky: using periastron frequencies as the initial condition when generating waveform templates and calculating the SNR. This refined approach will maximize the scientific output from future GW observations of eccentric binary systems. This, in turn, will enable more precise multi-messenger follow-ups, allowing us to connect GW signals with electromagnetic and other astronomical observations, offering more profound insights into the origin of these systems, their post-merger dynamics, gamma-ray burst generation, remnant properties, and signatures in the electromagnetic spectrum.

Furthermore, we incorporated subdominant modes alongside the dominant $(2,2)$ mode and analyzed the results across a range of binary parameters. Our findings indicate that including subdominant modes helps reduce the sky area for eccentric binaries, similar to the circular case. Notably, with increasing eccentricity, the sky area decreases even further compared to the circular case, both with and without the inclusion of subdominant modes. Specifically, for a binary system with masses $m_1 = 10 \, M_\odot$ and $m_2 = 1.4 \, M_\odot$ at an eccentricity of $e = 0.4$, the extra early warning time gained solely by including the subdominant modes is $27$ seconds, $54$ seconds, and $11.6$ minutes for O5, Voyager, and 3G scenarios at a sky area of 1000 sq. deg., respectively. We explored a parameter space with primary mass $m_1 \in [1.4, 15] \, M_\odot$, dimensionless spin $\chi_1 \in [0, 0.8]$, and eccentricity $e \leq 0.4$ for various detector sensitivities. We quantified the fractional reduction in sky area with eccentricity compared to circular binaries, both with and without subdominant modes. For NSBH binaries, we observed a sky area reduction of 2\% to 80\% as eccentricity increases from $e_5 = 0.1$ to $e_5 = 0.4$, with an added 25 seconds early warning for the dominant mode in the O5 scenario. BNS binaries showed a $40 \%$ reduction at $e = 0.4$, with even greater reductions when subdominant modes are considered.

Specific improvements with subdominant mode for different detector scenarios are as follows:

\textbf{O5:} The fractional reduction in sky area due to eccentricity for the $(2,2)$ mode plus subdominant modes can range from 2\% to 80\% as the eccentricity increases from $e_5 = 0.1$ to $e_5 = 0.4$ at a sky area of 1000 sq. deg., providing an extra early warning time of $41$ s. At a sky area of 100 sq. deg., this reduction can vary from 2\% to 40\% with increasing eccentricity. Furthermore, we can achieve a $70\%$ sky area reduction due to the inclusion of subdominant modes for NSBH systems, whereas for BNS systems, this reduction is near zero.

\textbf{Voyager:} The fractional reduction in sky area due to eccentricity can vary from 2.2\% to 85\% as the eccentricity increases from 0.1 to 0.4 at a sky area of 1000 sq. deg., resulting in an extra warning time of more than 1 minute. For a sky area of 100 sq. deg., this reduction can vary from 2\% to 65\%. The sky area reduction due to subdominant modes can reach up to $94\%$ for NSBH binaries, while for BNS binaries, this reduction is close to zero.

\textbf{3G:} For the 3G scenario, considering subdominant mode along with dominant mode, the reduction in sky area due to eccentricity can reach up to 80\% at early warning time of 9.5 minutes before merger at a sky area of 100 sq. deg. with an eccentricity of 0.4 for NSBH systems. For BNS systems, the fractional reduction in sky area can reach up to 60\% at 100 sq. deg. with an early warning time of 16 minutes at an eccentricity of 0.4. Furthermore, at a larger sky area of 1000 sq. deg., the early warning time from such binaries can extend to 40 minutes. The inclusion of subdominant modes can lead to a sky area reduction of up to 98\% for NSBH binaries during the inspiral phase. For BNS binaries, a reduction of less than 20\% can be achieved during the early inspiral; however, this reduction diminishes as the merger approaches.

Additionally, we explored the effect of component spins on the sky area for the O5 detector scenario, and found that the sky area weakly depends on spin, with the reduction in sky area due to spin being less than 12\% across all considered eccentricities.

We conclude that the template starting point is crucial for the study of eccentric gravitational-wave signals, particularly in low-latency analysis. Our research shows that templates beginning at periastron frequencies offer optimal SNR and sky localization for these signals. Therefore, in future studies, we plan to incorporate frequency and eccentricity into the template bank study, fixing the mean anomaly at periastron for eccentric signals.

\begin{acknowledgments}
We thank Ajith Parameswaran, Chandra Kant Mishra, Bala Iyer and the relativistic astrophysics research group at the International Centre for Theoretical Sciences (ICTS) for insightful discussions related to this work. 
P.S., P.R., A.M. and P.K. acknowledge support of the Department of Atomic Energy, Government of India, under project no. RTI4001. P.S. and P.K. also acknowledge support by the Ashok and Gita Vaish Early Career Faculty Fellowship at ICTS. Computational work for this study was carried out using the Alice and Sonic clusters at ICTS.
\end{acknowledgments}

\bibliography{main}

\end{document}